\documentclass[aps,prd,superscriptaddress,nofootinbib,showpacs,twocolumn]{revtex4}
\usepackage{amsmath,amssymb}
\usepackage{epsfig,graphicx}
\usepackage{color}
\usepackage{nccmath}

\catcode`@=11
\@addtoreset{equation}{section}
\catcode`@=12

\begin{document}
\title{Dynamics of the generalized unimodular gravity theory}
\author{A. O. Barvinsky}
\email{barvin@td.lpi.ru}
\affiliation{Theory Department, Lebedev
Physics Institute,
Leninsky Prospect 53, Moscow 119991, Russia}

\author{N.Kolganov}
\email{nikita.kolganov@phystech.edu}
\affiliation{Moscow Institute of Physics and Technology, Institutskii per. 9, Dolgoprudny 141700, Russia}
\affiliation{Bogoliubov Laboratory of Theoretical Physics, JINR, Joliot-Curie 6, Dubna 141980, Russia}

\author{A.Kurov}
\email{kurov.aleksandr@physics.msu.ru}
\affiliation{Theory Department, Lebedev Physics Institute, Leninsky Prospect 53, Moscow 119991, Russia}

\author{D.Nesterov}
\email{nesterov@td.lpi.ru}
\affiliation{Theory Department, Lebedev Physics Institute, Leninsky Prospect 53, Moscow 119991, Russia}


\begin{abstract}
The Hamiltonian formalism of the generalized unimodular gravity theory, which was recently suggested as a model of dark energy, is shown to be a complicated example of constrained dynamical system. The set of its canonical constraints has a bifurcation -- splitting of the theory into two branches differing by the number and type of these constraints, one of the branches effectively describing a gravitating perfect fluid with the time-dependent equation of state, which can potentially play the role of dark energy in cosmology. The first class constraints in this branch generate local gauge symmetries of the Lagrangian action -- two spatial diffeomorphisms -- and rule out the temporal diffeomorphism which does not have a realization in the form of the canonical transformation on phase space of the theory and turns out to be either nonlocal in time or violating boundary conditions at spatial infinity. As a consequence, the Hamiltonian reduction of the model enlarges its physical sector from two general relativistic modes to three degrees of freedom including the scalar graviton. This scalar mode is free from ghost and gradient instabilities on the Friedmann background in a wide class of models subject to a certain restriction on time-dependent parameter $w$ of the dark fluid equation of state, $p=w\varepsilon$. For a special family of models this scalar mode can be ruled out even below the phantom divide line $w=-1$, but this line cannot be crossed in the course of the cosmological expansion. This is likely to disable the generalized unimodular gravity as a model of the phenomenologically consistent dark energy scenario, but opens the prospects in inflation theory with a scalar graviton playing the role of inflaton.
\end{abstract}

\maketitle

\section{Introduction}
Recently suggested theory of the generalized unimodular gravity (GUMG) \cite{darkness} was motivated by the necessity to build a model of dark energy with a variable in time equation of state that could fit cosmological acceleration data. As a candidate for dark energy, this model oversteps the limitations of the simplest models -- Einstein general relativity (GR) with a fundamental cosmological constant and the unimodular gravity (UMG) theory \cite{UMG}. In GR the global degree of freedom responsible for cosmological acceleration exists in the form of the fundamental cosmological constant $\varLambda$, while the UMG theory is more flexible because $\varLambda$ arises as an integration constant of equations of motion and should be fixed by the choice of initial conditions. This flexibility is not sufficient, however, to unleash dark energy evolution, $\dot\varLambda\neq 0$. In particular, it does not allow the system to cross the phantom divide line $w=-1$ of the equation of state parameter \cite{PDL,Vikman}, which is likely to be indicated by observations \cite{DE_data,DE_data1}.

Quite interestingly, this dilemma of obtaining non-clustering (that is global or homogeneous in space) but evolving in time physical mode can be solved by generalizing the unimodular gravity theory -- replacing the UMG condition of a unit determinant of the spacetime metric $g_{\mu\nu}$ by the following condition
    \begin{equation}
    (-g^{00})^{-1/2}=N(\gamma),
    \quad\gamma=\det \gamma_{ij},               \label{condition}
    \end{equation}
with some rather generic function $N(\gamma)$ of the determinant of the 3-dimensional metric $\gamma_{ij}$. Using this kinematical restriction on metric coefficients in the Einstein-Hilbert action $S_{EH}[\,g_{\mu\nu}]$ as the definition of the GUMG action,
    \begin{equation}
    S_{GUMG}[\,g_{ij}, g_{0i}\,]=S_{EH}[\,g_{\mu\nu}]\,
    \big|_{\;(-g^{00})^{-1/2}=N(\gamma)}\,\,,            \label{GUMGaction}
    \end{equation}
one obtains the theory whose equations of motion effectively coincide with Einstein equations in the presence of a perfect fluid which has a barotropic equation of state $p=w\varepsilon$, its parameter $w=w(\gamma)$ being determined by the function $N(\gamma)$ and therefore evolving in time \cite{darkness}. Important feature of the model is that this dark fluid is built of purely gravitational degrees of freedom -- dark energy constituent of the theory is not achieved by adding extra fields, like for example quintessence \cite{quintessence}, Chaplygin gas \cite{Chaplygin}, dilaton \cite{dilaton} or khronon \cite{khronon}, but composed of the original spacetime metric variables. As a modification of Einstein gravity, it also does not fall into the category of higher-derivative \cite{Starobinsky_model} or nonlocal gravity \cite{nonlocal} models, because it is based on an ultralocal kinematical restriction on the lapse function of the theory.

An obvious advantage of this model is that it should not actually be required to pass typical tests on correspondence principle with GR phenomenology, because every solution of Einstein theory solves GUMG equations of motion -- as one can show, nine GUMG equations of motion are just the projections of ten Einstein equations onto the subspace of metric coefficients satisfying the condition (\ref{condition}). Reversed statement is not correct -- there are nontrivial solutions in GUMG theory, which are absent in general relativity. Equivalently, GUMG theory generically has additional physical degrees of freedom missing in GR. Their dynamics might lead to classical and quantum instabilities including as a particular case clustering phenomenon which is not acceptable for dark energy constituents. The attempt of addressing this problem was recently undertaken in \cite{Bufalo_et_al}, but it is hard to agree with various points and conclusions of this work. Therefore, the main goal of our paper is to make a systematic analysis of the physical degrees of freedom in GUMG theory and find out how robust is this theory against potential ghost and gradient instabilities.

The dynamical content of the physical sector strongly depends on local gauge symmetries and their realization as canonical transformations in the constrained Hamiltonian formalism of the theory. Therefore, we develop here the canonical formalism of GUMG model and immediately find out that it is very interesting from the viewpoint of the theory of constrained dynamics \cite{Henneaux-Teitelboim}. It turns out to be essentially more complicated than that of the Einstein GR or UMG. It has several generations of constraints which in terminology of \cite{Henneaux-Teitelboim} bifurcate, that is split the theory into two branches differing both by the number of constraints and their type -- of the first and of the second class. Moreover, the first class constraints in the physically interesting branch, associated with the presence of dark fluid, do not separately belong to definite generations (primary, secondary, etc.) but represent nontrivial linear combinations of those. This requires to modify the known Henneaux-Teitelboim procedure of recovering the Lagrangian gauge symmetries from the canonical transformations generated by first class constraints \cite{Henneaux-Teitelboim}. We develop such a modification and find that the number of local gauge symmetries of the Lagrangian action is actually smaller than their number anticipated in \cite{darkness}, so that GUMG theory seems to violate the so-called Dirac conjecture of constrained dynamics -- equality of the number of Lagrangian gauge symmetries and the number of primary first class constraints. It turns out, however, that the Dirac conjecture still remains true, because one of the three spacetime diffeomorphisms of GUMG theory considered in \cite{darkness} is actually nonlocal in time (or violates boundary conditions at spatial infinity) and, therefore, does not participate in the Hamiltonian reduction to the physical sector. Ultimately this leads to the third local degree of freedom -- the scalar ``graviton" which is absent in GR and UMG theory.

We find the range of perturbative stability of the theory on the cosmological Friedmann background -- the class of functions $w(\gamma)$ for which this scalar graviton is free from ghost and gradient instabilities. With a dynamical scalar graviton this range implies $w>-1$. Moreover, there is a special family of models in which the scalar graviton is not dynamical, so that stability citeria essentially relax and this range extends below the phantom divide line $w=-1$. In any case, however, this line is not crossed in the course of the cosmological expansion, which makes GUMG model hardly feasible as a candidate for dark energy scenario. On the other hand, this model turns out to be interesting as a possible source of inflation, the scalar graviton playing the role of inflaton having a nontrivial speed of sound, which is briefly discussed in conclusions of the paper.

\section{Gauge invariance in the generalized unimodular gravity}
The GUMG theory action (\ref{GUMGaction}) can equivalently be rewritten in terms of ten independent metric coefficients with the kinematic relation (\ref{condition}) enforced via the Lagrange multiplier $\lambda$ \cite{darkness},\footnote{For simplicity of the formalism we use the units in which the gravitational coupling constant is dimensionaless and equals $G=1/16\pi$.}
    \begin{equation}
    \begin{aligned}
    &S_{GUMG}[\,g_{\mu\nu}, \lambda\,]=
    \int\limits_{[\,t_-,t_+]\times\varSigma} d^4x \, \Bigl[\, g^{1/2} R \\
    &\qquad\qquad- \lambda \bigl(\,(-g^{00})^{-\frac12} - N(\gamma)\, \bigr) \,\Bigr]+S_B.
    \label{GUMG_original}
    \end{aligned}
    \end{equation}
Here $S_B=S_\perp+S_{\,\vdash}$ is the boundary term of the Gibbons-Hawking type which effectively removes due to integration by parts second-order spacetime derivatives from the Einstein-Hilbert Lagrangian. It consists of the surface terms $S_\perp$ at the future and past spacelike boundaries $\varSigma_\pm$ at $t_\pm$ and the surface term $S_{\,\vdash}$ at the ``side" timelike boundary of the topology $[\,t_-,t_+]\times\partial\varSigma$, $\partial\varSigma$ being the boundary of spatial slices $\varSigma$ -- their structure will be discussed in more detail below.

Addition of these surface terms guarantees consistency of the variational procedure for this action subject to fixed induced metric of the boundary of the spacetime domain $[\,t_-,t_+]\times\varSigma$. As a result the action is stationary on the configurations satisfying the Einstein equations,
    \begin{align}
    &R_{\mu\nu} - \frac12\, g_{\mu\nu} R
    = \frac12\, T_{\mu\nu},              \label{Ee}\\
    &T_{\mu\nu} = \frac2{\sqrt{g}} \frac{\delta}{\delta g^{\mu\nu}} \int d^4 x \, \lambda \, \bigl((-g^{00})^{-1/2} - N(\gamma) \bigr)\nonumber\\
    &\qquad\qquad= (\varepsilon + p) \, u_\mu u_\nu + p \, g_{\mu\nu},
    \end{align}
with the stress tensor of the effective perfect fluid which has a 4-velocity $u_\mu =-\delta^0_\mu\,N$ and the equation of state with a variable barotropic parameter $w$,
    \begin{equation}
    p = w \varepsilon, \quad w
    = 2\,\frac{d \ln N(\gamma)}{d \ln \gamma},
    \quad \varepsilon = \frac{\lambda}{\sqrt{\gamma}}.  \label{w}
    \end{equation}
Nonconstant nature of $w$ is what distinguishes this model from  unimodular gravity (corresponding to $w=-1$) and serves as a main motivation to consider GUMG as a candidate for dark energy which has the phenomenological equation of state with $w$ essentially depending on the red shift parameter.

The energy density $\varepsilon$ here is in fact composed of the metric,
    \begin{equation}
    \varepsilon=2\,u^\mu u^ \nu \, G_{\mu\nu}, \quad
    G_{\mu\nu} = R_{\mu\nu} - \frac12\, g_{\mu\nu} \, R,    \label{epsilon}
    \end{equation}
as it follows from the contraction of (\ref{Ee}) with $u^\mu u^ \nu$. Therefore, the ten equations (\ref{Ee}) are not independent, but represent the projection of the vacuum Einstein equations on the set of nine independent equations
    \begin{equation}
    \begin{aligned}
    &P_{\mu\nu}^{\;\;\;\;\rho\sigma}G_{\rho\sigma} = 0,\\
    &P_{\mu\nu}^{\;\;\;\rho\sigma}\equiv\delta^\rho{}_{\!\!\!(\mu} \delta^\sigma{}_{\!\!\!\nu)} - \big[\, u_\mu u_\nu + w ( u_\mu u_\nu + g_{\mu\nu})\, \big]\, u^\rho u^\sigma,
    \end{aligned}
    \end{equation}
as it should be for nine independent metric coefficients $g_{ij}$ and $g_{0i}$. Here the projector $P_{\mu\nu}^{\;\;\;\rho\sigma}$ has left and right zero eigenvalue eigenvectors, $u^\mu u^\nu P_{\mu\nu}^{\;\;\;\;\rho\sigma}=0$, $P_{\mu\nu}^{\;\;\;\rho\sigma} [\,u_\rho u_\sigma + w ( u_\rho u_\sigma + g_{\rho\sigma})\,]=0$.\footnote{This explains why every solution of vacuum Einstein theory is also a solution of the GUMG model, as mentioned in Introduction. The same property also holds for Einstein and GUMG theories with matter sources.}

The canonical formalism of the theory is usually described in terms of the ADM (3+1)-decomposition of the spacetime metric into the lapse function $N$, shift functions $N_i$ and the spatial metric $\gamma_{ij}$ of spacelike slices of a constant time $x^0=t$ \cite{ADM}
    \begin{equation}
    \begin{aligned}
    &N = (-g^{00})^{-1/2}, \quad N_i = g_{0i},
    \quad N^i = \gamma^{ij} N_j, \\
    &\gamma_{ij} = g_{ij}, \quad \gamma^{ij}
    = (\gamma_{ij})^{-1} = g^{ij} - \frac{N^i N^j}{N^2}.
    \end{aligned}     \label{ADMvariables}
    \end{equation}
In terms of these variables subtraction of Gibbons-Hawking boundary terms $S_\perp$ at the future and past spatial surfaces $\varSigma_\pm$ leads to the the action
    \begin{align}
    &S_{GUMG}[\,\gamma_{ij}, N^i\,]\nonumber\\
    &\,\,= \int\limits_{t_-}^{t_+} dt \int\limits_\varSigma d^3x \, N \sqrt{\gamma} \,\bigl(\, {}^3 \! R + K_{ij}^2 - K^2\,\bigr)\;\Big|_{\,N=N(\gamma)}+S_{\,\vdash},
    \label{GUMG_ADM_fixed}
    \end{align}
where the bulk term is the ADM action in terms the extrinsic curvature of the spatial slices
    \begin{equation}
    K_{ij} = \frac1{2N} (\nabla_i N_j + \nabla_j N_i
    - \dot \gamma_{ij}), \quad K = \gamma^{ij} K_{ij}, \label{K_def}
    \end{equation}
including spatial covariant derivatives $\nabla_i$  with respect to the 3-metric $\gamma_{ij}$ and the 3-dimensional scalar curvature ${}^3 \! R$ of this metric. The lapse function is explicitly expressed as a function of $\gamma$, which turns out to be more convenient than introducing this dependence via the Lagrange multiplier.

We will not give here a concrete form of the ``side" surface term $S_{\,\vdash}$. It certainly identically vanishes for spatially closed cosmology, and we will also explicitly describe its effect for the linearized theory on the asymptotically flat spacetime, where $S_{\,\vdash}$ is responsible for the ADM energy \cite{ADM}, and asymptotically Friedmann Universe. As we will see, boundary conditions critically influence gauge invariance properties of the GUMG theory, and the mechanisms of this influence are rather different in these two topologically closed and open cases.

Because of the GUMG kinematical restriction on metric coefficients (\ref{condition}) the theory is not invariant under local spacetime diffeomorphisms with a generic vector field parameter $\xi^\mu(t,x)$
    \begin{equation}
    \delta^\xi g_{\mu\nu}=D_\mu\xi_\nu+D_\nu\xi_\mu,\quad \xi_\mu=g_{\mu\nu}\xi^\nu,
    \end{equation}
($D_\mu$ is the covariant derivative with respect to the metric $g_{\mu\nu}$). As it was assumed in \cite{darkness}, because of general coordinate invariance of the Einstein action the GUMG action (\ref{GUMGaction}) at least naively is invariant under the subset of diffeomorphisms which preserve the condition (\ref{condition}). Rewritten in the basis of ADM variables (\ref{ADMvariables}) the diffeomorphisms with a generic $\xi^\mu = (\xi^0, \xi^i)$ read
    \begin{align}
    &\delta^\xi N= N \dot \xi^0 + \dot N \xi^0
    - N N^l \partial_l \xi^0 + \xi^l\partial_l N, \label{lapse_diffeo}\\
    &\delta^\xi N^i= N^i \dot \xi^0 + \dot N^i \xi^0
    - (N^2 \gamma^{ij} + N^i N^j)\, \partial_j \xi^0\nonumber\\
    &\qquad\quad+ \dot \xi^i+ \xi^l\partial_l N^i
    - N^l \partial_l \xi^i,                  \label{shift_diffeo}\\
    &\delta^\xi \gamma_{ij}=
    \dot \gamma_{ij} \xi^0 + 2 N^l \gamma_{l(i}
    \partial_{j)}\xi^0 + \xi^l \partial_l\gamma_{ij}
    + 2 \gamma_{l(i} \partial_{j)} \xi^l,          \label{gamma_diffeo}
    \end{align}
so that the preservation of the condition $N-N(\gamma)=0$ reduces to the differential equation on $\xi^\mu$
    \begin{equation}
    \delta^\xi \bigl[\,N - N(\gamma)\,\bigr] =
    N\bigl[\,\partial_t \xi^0
    - (1 + w) N^k \partial_k \xi^0
    - w \, \partial_l \xi^l\,\bigr] = 0.  \label{xi_equation}
    \end{equation}

From the viewpoint of the dynamical content of the theory, the invariance transformations of the action can be considered as gauge ones (that is not changing the physical state of the system) only if they are local in time, which means that they should be labelled by independent gauge parameters and their time derivatives to some finite order. Then the invariance transformations can be endowed with a finite support within the full time range $[\,t_-,t_+]$ and should vanish together with all their time derivatives at $t_\pm$, thus changing fields only inside the spacetime sandwich. This means that the solution of (\ref{xi_equation}) for $\xi^\mu(x)$ should not contain nonlocality in time. At least naively, this does not apply to nonlocal properties of this solution in space -- spatially nonlocal but local in time system can be considered as a legitimate gauge constrained system subject to canonical quantization \cite{Henneaux-Teitelboim}. This means that Eq.(\ref{xi_equation}) should be solved for the spatial components $\xi^i$ of $\xi^\mu$ in terms of its temporal one $\xi^0$ and its time derivative, but not vice versa. There are two solutions of this type, which were suggested in \cite{darkness} -- one is given by a purely spatial 3D transversal vector
    \begin{equation}
    \xi^\mu = \begin{bmatrix}
    0 \\
    \\
    \;\xi_\bot^i\,
    \end{bmatrix}, \quad \partial_k \xi_\bot^k = 0, \label{xi_trans}
    \end{equation}
and the second one has a 3D longitudinal component parameterized by a nonvanishing $\xi^0$ and $\dot\xi^0$
    \begin{equation}
    \xi^\mu =
    \begin{bmatrix}
    \xi^0 \\
        \\
    \;\partial^i \frac1{\Delta}\frac{{\mathcal D}_t \xi^0}w
    \end{bmatrix}, \quad
    {\mathcal D}_t  = \partial_t
    - (1 + w) N^k \partial_k,                      \label{xi_long}
    \end{equation}
where $1/\Delta$ is the nonlocal operation -- the inverse of the spatial Laplacian $\Delta = \partial^i\partial_i$, $\partial^i\equiv\delta^{ij}\partial_j$. Here important subtlety arises which distinguishes the case of spatially closed models without a boundary from the case of asymptotically flat spacetimes or Friedmann cosmologies with asymptotically flat spatial slices.

\subsection{Spatially closed models}
Closed GUMG models are important for the theory of cosmological perturbations on the background of the Friedman metric with the $S^3$-topology of the spatial slices $\varSigma$. To begin with, we will modify the formulation of GUMG theories -- incorporate a kind of 3-dimensional bimetric covariance by introducing in the condition (\ref{condition}) the dependence on auxiliary spatial metric $\sigma_{ij}$,
    \begin{equation}
    N(\gamma)\to N(\gamma/\sigma),\quad
    \sigma={\rm det}\,\sigma_{ij}.              \label{comp_mod}
    \end{equation}
Then this condition becomes a scalar with respect to {\em simultaneous} coordinate transformations of two spatial metrics $\gamma_{ij}$ and $\sigma_{ij}$ and allows one to consider the model in an arbitrary coordinate system on $\varSigma$.\footnote{This also allows one to avoid coordinate singularities, which in the absence of this additional metric would necessarily mar the formalism. Say, in the case of $\varSigma=S^3$ in the natural spherical coordinate system the determinant $\gamma$ vanishes at the poles of the 3-sphere and becomes not invertible, whereas with the choice of $\sigma_{ij}$ as a metric of the unit 3-sphere the ratio $\gamma/\sigma$ becomes regular throughout the whole $S^3$. We thank A.Kulyabin for this observation.} The auxiliary metric can be taken time independent, but generally has curvature and involves spatial coordinates, $\sigma_{ij}=\sigma_{ij}({x})$, ${x}=x^i$. Then Eq.(\ref{xi_equation}) gets modified by a simple replacement of all 3-dimensional partial derivatives with covariant derivatives for the metric $\sigma_{ij}$,
    \begin{equation}
    \partial_i\to\bar\nabla_i,\quad\bar\nabla_k\sigma_{ij}=0,\quad
    \bar\nabla^i=\sigma^{ij}\bar\nabla_j.
    \end{equation}
Note that the transformations (\ref{lapse_diffeo})-(\ref{gamma_diffeo}) can also be rewritten by this replacement in the bimetric covariant form which involves nonvanishing $\bar\nabla$-derivatives of the dynamical metric $\gamma_{ij}$.\footnote{It should be emphasized that the {\em bimetric} covariance does not bring in the theory local gauge invariance because an auxiliary metric is not dynamical -- it plays in the action the role of external parameter which is not subject to variations in the variational principle.}

The same rule applies to the solutions (\ref{xi_trans}) and (\ref{xi_long}), but there is a problem with the latter. For the nonlocal longitudinal vector
    \begin{equation}
    \xi^i_\| =\bar\nabla^i \frac1{\bar\Delta}
    \frac{{\mathcal D}_t \xi^0}w           \label{xi_long1}
    \end{equation}
to be well defined the function ${\mathcal D}_t \xi^0/w$ should not contain a constant zero mode of the covariant Laplacian $\bar\Delta=\sigma^{ij}\bar\nabla_i\bar\nabla_j$. On closed compact $\varSigma$ this scalar Laplacian has a discreet spectrum with a single constant mode, and this condition reduces to
    \begin{equation}
    \int_\varSigma d^3x\,\sqrt\sigma\,\frac{\dot\xi^0
    - (1 + w) N^k \partial_k\xi^0}w=0.
    \end{equation}
This is in complete agreement with the fact that the covariant version of the equation (\ref{xi_equation}) implies on compact space without a boundary that
    \begin{equation}
    \int_\varSigma d^3x\,\sqrt\sigma\,\frac{\dot\xi^0- (1 + w) N^k \partial_k \xi^0}w=
    \int_\varSigma d^3x\sqrt\sigma\,
    \bar\nabla_l\xi^l\equiv 0.                   \label{xi_equation1}
    \end{equation}

Therefore, $\xi^0(t,{x})$ should satisfy a nontrivial differential in time equation, so that it cannot have a compact support within the time range and cannot be treated as a local gauge parameter. Thus, on closed compact space $\xi^0$-transformation is not a gauge transformation of the theory -- it nontrivially changes field configurations at $t_\pm$ and corresponds to a change in the physical state of the system. Only the 3-dimensional transverse vectors (\ref{xi_trans}) generate local gauge transformations on a compact closed space\footnote{Transverse 3-dimensional vectors $\xi^i_\perp$ can be rendered local compact support in the overcomplete basis of dual two-forms by choosing the representation $\xi^i_\perp=\epsilon^{ijk}\bar\nabla_j\lambda_k$ in terms of an arbitrary covector field $\lambda_k$ with a compact support.}  -- the property that was not envisaged in \cite{darkness}. Below we will confirm this property both in the canonical formalism of the full nonlinear GUMG theory and by explicit calculations in the linearized theory on the homogeneous Friedmann background.

\subsection{Asymptotically flat models}
In asymptotically flat models breakdown of gauge invariance under $\xi^0$-transformations has a different mechanism. In this case the most natural choice of the auxiliary metric is $\sigma_{ij}=\delta_{ij}$, the flat space Laplacian has a continuous spectrum and the kernel of its Green's function
    \begin{equation}
    \frac1{\bar\Delta}\,\delta(x-y)=
    -\frac1{4\pi\,|\,x-y\,|}\sim \frac1{|\,{x}\,|},\;\;\;
    |\,{x}\,|\to\infty,
    \end{equation}
generates at the spatial infinity the multipole expansion for $\xi^i_\|$ in (\ref{xi_long1}), beginning with
    \begin{equation}
    \xi^i_\|(x)\sim \frac1{|\,{x}\,|^2}.
    \end{equation}
In contrast to the closed model case no equations imposed on $\xi^0$ are needed for the existence of this expansion, except the requirement of the integration convergence in the convolution of the Green's function kernel and ${\mathcal D}_t \xi^0(y)$. This can be attained by imposing the falloff condition at infinity ${\mathcal D}_t \xi^0(y)\sim 1/|\,y\,|^3$ which is obviously guaranteed for $\xi^0(x)$ with a compact support. In contrast to $\xi^0(x)$, however, $\xi^i_\|(x)$ slowly falls off at infinity, so that the surface integral over the remote spacetime boundary turns out to be finite and nonvanishing,
    \begin{equation}
    \int dt\!\!\int\limits_{|x|\to\infty}d^2\varSigma_i(x)\;
    \xi^i_\|(x)\neq 0.                                 \label{asymp_int}
    \end{equation}
As we show below in the linearized theory, exactly this integral contributes to the gauge transformation of the GUMG action and breaks its invariance. Thus, $\xi^0$-transformation also does not belong to local gauge symmetries of the GUMG theory.

\section{Bifurcation of canonical constraints}
Canonical formalism for the Lagrangian action (\ref{GUMG_ADM_fixed}) begins with the definition of momenta conjugated to phase space coordinates $N^i$ and $\gamma_{ij}$
\begin{align}
&P_i= 0, \label{P_def} \\
&\pi^{ij}= - \sqrt{\gamma} (K^{ij} - \gamma^{ij} K), \label{pi_def}
\end{align}
and the Legendre transform with respect to phase space velocities from the Lagrangian to the Hamiltonian $\cal H$. Thus we get the primary constraint (\ref{P_def}), and the Hamiltonian takes the form of the linear combination of what is called in Einstein theory the Hamiltonian $H_\perp$ and momenta $H_i$ constraints,
    \begin{align}
    &{\cal H} = \int_\varSigma d^3x \, \bigl(N H_\bot
    + N^i H_i \bigr)+{\cal H}_{\;\vdash},            \label{Hamiltonian}\\
    &H_\bot=\!\frac{\gamma_{in}\gamma_{jm}\pi^{ij}\pi^{mn}\! - \frac12\,\pi^2}{\sqrt\gamma}\!-\!\sqrt{\gamma} \, {}^3\! R,\,\,
    \pi=\gamma_{ij}\pi^{ij}, \\
    &H_i= -2\gamma_{ij}\nabla_k \pi^{jk}.
    \end{align}
The Hamiltonian also includes a spatial surface term ${\cal H}_{\;\vdash}$ inherited from integrations by parts and the surface term in the action (\ref{GUMG_ADM_fixed}), but here we will not go in its details and disregard it, though of course it plays important role by making the variational procedure consistent under fixed boundary conditions at $\partial\varSigma$.

As a result, the theory can be equivalently reformulated as a variational principle for the total canonical action including the primary constraints with the Lagrange multipliers $u_1^i$
    \begin{align}
    &S_T[\,\gamma_{ij},\pi^{ij},N^i,P_i,u_1^i\,] = \int dt \, d^3x \, \bigl(\pi^{ij} \dot \gamma_{ij} + P_i \dot N^i\nonumber\\
    &\qquad\qquad\qquad\qquad- N H_\bot - N^i H_i
    - u_1^i P_i \bigr),                    \label{action_T}
    \end{align}
In the conventional terminology \cite{Henneaux-Teitelboim} this is the {\em total} action of the constrained system. For all phase space functions $\varPhi$ it generates their canonical evolution via the Poisson bracket $\{...,...\}$ defined on the space of canonical coordinates and momenta,
    \begin{equation}
    \dot\varPhi=\big\{\,\varPhi,{\cal H}+{\textstyle\,\int} d^3 y \, u_1^k(y) P_k(y)\,\big\},
    \end{equation}
and yields the primary constraints (\ref{P_def}) by varying their Lagrange multipliers.

Unlike in Einstein theory, $H_\bot$ here is no longer a constraint because the lapse function is not an independent variable, but the momenta constraints
    \begin{equation}
    H_i = 0.
    \end{equation}
arise as secondary constraints from the requirement of conservation of the primary ones, $\dot P_i=0$, because $\{P_i,{\cal H}+{\textstyle\int} d^3 y \, u_1^k(y) P_k(y)\}=-H_i$. In its turn, their conservation leads in view of the Poisson bracket commutator $\{H_i,{\cal H}\} =\partial_k (N^k H_i)+H_k \partial_i N^k+\partial_i(w N H_\bot )$ to the new tertiary constraints
    \begin{equation}
    T_i=0,\quad
    T_i = \partial_i T,\quad T\equiv
    w N H_\bot, \label{tertiary}
    \end{equation}
where $w=w(\gamma)$ is given by (\ref{w}). Conservation of $T_i$ results in the following relation
    \begin{equation}
    \{T_i,{\cal H}\} = \partial_i \Bigl(w \, \partial_k \bigl( N^2 \gamma^{kl} H_l \bigr) + N^k T_k + T \, S \Bigr) = 0,\label{T_stability}
    \end{equation}
where
    \begin{align}
    &S =\varOmega\,\partial_k N^k
    - \frac{d \ln w}{d \ln \gamma}
    \frac{\pi N}{\sqrt{\gamma}},            \label{S}\\
    &\varOmega= 1 + w + 2 \frac{d \ln w}{d \ln \gamma}.   \label{Omega}
    \end{align}
Omitting in (\ref{T_stability}) the terms which are already proportional to the existing constraints, one can rewrite (\ref{T_stability}) as $T \, \partial_i S = 0$. This immediately leads to two possible quaternary constraints:
    \begin{equation}
    T= 0
    \end{equation}
(which is equivalent to general relativistic constraint $H_\perp=0$) or
    \begin{equation}
    \partial_i S = 0.
    \end{equation}
In the terminology of \cite{Henneaux-Teitelboim} this is a bifurcation of the system of constraints -- the theory has two dynamically different branches corresponding to these cases. No further constraints appear in both of these branches. Indeed, in the first branch the conservation of $T$
    \begin{equation}
    \{\,T, {\cal H}\,\} = w \, \partial_k \bigl( N^2\gamma^{kl} H_l \bigr)+N^k T_k+T \, S = 0
    \end{equation}
is proportional to the existing constraints, while in the second branch the conservation of $S_i$ is just an equation on the Lagrange multiplier $u_1^i$.

All secondary and higher order constraints are the corollaries of dynamical equations. To see this as a byproduct and get the Lagrangian expressions for constraints we note that in view of (\ref{Ee}) the effective stress tensor of the perfect fluid is covariantly conserved, $\nabla^\mu T_{\mu\nu}=0$. With $u^\mu=-g^{0\mu}N$ the spatial component of this conservation law reads 
    \begin{equation}
    \nabla^\mu T_{\mu i} = \frac1{N\sqrt{\gamma}}
    \partial_i (\sqrt{\gamma} \varepsilon w N) = 0
    \end{equation}
But from (\ref{w}) and (\ref{epsilon}) it follows that $\varepsilon\sqrt\gamma=- H_\bot$, which immediately brings us to the tertiary constraint (\ref{tertiary}). On the other hand, the temporal component of Bianchi identities gives
    \begin{equation}
    u^\nu \nabla^\mu T_{\mu\nu} = \frac{\varepsilon}N \Bigl[ (1 + w) \partial_k N^k - (\partial_t - N^k \partial_k) \ln{\sqrt\gamma \varepsilon N} \Bigr] = 0,
    \end{equation}
where we took into account that $u^0=1/N$, $u^i=-N^i/N$, $p=w\varepsilon$ and the expression for $w$, $w=2d\ln N/d\ln\gamma$. If $\varepsilon=0$, then we get the $T=-w\varepsilon\sqrt\gamma N=0$ branch of the theory. If $T\neq 0$, then one can express the time derivative of $\sqrt\gamma\,\varepsilon N$ from the equation above and use it in the conservation law $\dot T_i=-\partial_i\partial_t(\sqrt{\gamma} \varepsilon w N) = 0$, which gives $\dot T_i = T\, S_i = 0$, where
    \begin{equation}
    S_i = \partial_i\bigl[(1 + w) \, \partial_k N^k + (\partial_t - N^k \partial_k) \ln w \bigr]
    \end{equation}
coincides with the Lagrangian form of the constraint $S_i$ -- this can be directly verified by substituting into (\ref{S}) the Lagrangian expressions for momenta. Therefore, with $T\neq 0$ we recover the $S_i = 0$ branch.

The last comment of this section concerns the compact space modification (\ref{comp_mod}). Transition to this case in the above formulae with the {\em time-independent} metric $\sigma_{ij}$ consists, as was mentioned above, in the replacement of all partial derivatives by covariant derivatives with the Riemannian $\sigma$-metric connection, acting on relevant tensors or tensor densities. In particular, since $H_\perp$ is a scalar density, the covariant derivative $\bar\nabla_i$, which should replace $\partial_i$ in the constraint $T_i$, reads as
    \begin{equation}
    T_i = \bar\nabla_i T\equiv
    \sqrt\sigma\,\partial_i\frac{w N H_\bot}{\sqrt\sigma}. \label{tertiary1}
    \end{equation}

\section{Algebra of constraints for two branches of GUMG theory}
As we saw, the sets of constraints are different for two bifurcating branches. Here we consider their Poisson bracket algebras and begin with the $T=0$ case. In this case we have primary $P_i$, secondary $H_i$ and tertiary $T$ constraints which all belong in the Dirac terminology to the first class, because their commutators with each other are vanishing on their full constraint surface $(P_i,H_i,T)=0$,
    \begin{align}
    &\!\!\!\{P_i(x), P_j(y)\}\!=\! \{P_i(x), H_j(y)\}\!
    =\! \{P_i(x), T(y)\}\! =\! 0,                    \label{PB_primary}\\
    &\!\!\!\{H_i(x), H_j(y)\}= H_j(x)\,\partial_i\delta(x, y)
    - (i, x \leftrightarrow j, y),                      \label{PB_M_M}\\
    &\!\!\!\{H_i(x), T(y)\}=\!- \delta(x, y) \,
    \partial_i T\! - \!\big(\varOmega\, T\big)(y)\,
    \partial_i \delta(y,x) , \\
    &\!\!\!\{T(x), T(y) \}\nonumber\\
    &\quad=\big(wN\big)(y)\,\big(\gamma^{ij}wNH_i\big)(x)\,
    \partial_j \delta(x,y)- (x\!\leftrightarrow\! y), \label{PB_T_T}
    \end{align}
and they also commute with the Hamiltonian
    \begin{align}
    &\{ {\cal H}, P_i \}= H_i,                 \label{PB_H_P} \\
    &\{ {\cal H}, H_i \}= - \partial_k (N^k H_i)
    - H_k \partial_i N^k - \partial_i T, \\
    &\{{\cal H}, T\}= - w \, \partial_k \bigl(\,N^2 \gamma^{kl} H_l \bigr) - N^k \partial_k T - T \, S. \label{PB_H_T}
    \end{align}
For this branch of the theory the stress tensor of the effective perfect fluid is obviously vanishing because $\varepsilon=-T/w\sqrt\gamma N=0$, and the equations of motion coincide with the vacuum Einstein equations. Obvious interpretation of this GUMG branch is that it is a partial gauge fixing of the Einstein theory in the gauge (\ref{condition}).

Interestingly, the algebra of constraints in the $T=0$ branch of the theory is very similar to the case of the unimodular gravity, even though in UMG with $w = -1$ and $N(\gamma) = 1/ \sqrt{\gamma}$, the function $T=-H_\perp/\sqrt\gamma$ is no longer a constraint, but rather a constant of motion fixed by initial conditions. Indeed, in this case $\varOmega\equiv 0$ and the $S$-constraint (\ref{S}) is absent, so that $P_i$, $H_i$ and $T_i$ form the full set of constraints. Moreover, their Poisson bracket algebra is closed -- the first two sets of commutators coincide with (\ref{PB_primary}) and (\ref{PB_M_M}), while the rest of them read
    \begin{align}
    &\big\{H_i(x), T_j(y)\big\}= - \partial_j \bigl( T_i\,
    \delta(y, x)\, \bigr), \\
    &\big\{T_i(x), T_j(y) \big\}=\!\partial^x_i\partial^y_j \bigl(N(y)\big(\gamma^{kl}N
    H_k\big)(x)\,\partial_l \delta(x, y)\bigr)\nonumber\\
    &\qquad\qquad\qquad\quad- (i,x \leftrightarrow j,y).
    \end{align}
These constraints also commute with the Hamiltonian as (\ref{PB_H_P}) and
    \begin{align}
    &\{ {\cal H}, H_i \}= - \partial_k \bigl(N^k H_i\bigr)
    - H_k \partial_i N^k - T_i, \\
    &\{{\cal H}, T_i\}=\partial_i
    \bigl(\partial_k (N^2 \gamma^{kl} H_l)
    - N^k T_k \bigr).                              \label{PB_H_T_um}
    \end{align}
Thus, in unimodular gravity, just like in general relativity, all constraints belong to the first class \cite{UMG,UMG2,Ohta_et_al}.

The second branch of $T\neq 0$ and $S_i=0$ is nontrivial and physically much more interesting because it incorporates the effective perfect fluid simulating the role of dark energy. It has four generations of constraints from primary to quaternary ones for which we will use the collective notation
    \begin{equation}
    \phi_I = \bigl( P_i, H_i, T_i, S_i \bigr).     \label{all_constr}
    \end{equation}
Their Poisson bracket commutators form the matrix
\begin{widetext}
\begin{equation}
\{ \phi_I(x), \phi_J(y) \} =
\begin{bmatrix}
0 & 0 & 0 & \;\;\vphantom{\frac{\hat L}{\hat L}} \{P_i(x), S_j(y)\}\;\; \\
0 & 0 & \;\;\{H_i(x), T_j(y)\}\; & \vphantom{\frac{\hat L}{\hat L}} \{H_i(x), S_j(y)\} \\
0 & \vphantom{\frac{\hat L}{\hat L}} \;\;\{T_i(x), H_j(y)\}\; & 0 & \vphantom{\frac{\hat L}{\hat L}} \{T_i(x), S_j(y) \} \\
\;\;\{S_i(x), P_j(y) \} & \vphantom{\frac{\hat L}{\hat L}} \;\;\{S_i(x), H_j(y) \} & \{S_i(x), T_j(y) \} & 0
\end{bmatrix}\,.
\end{equation}
\end{widetext}
Its nonvanishing elements imply that the first class constraints, which commute with each other and with the Hamiltonian, should be disentangled from (\ref{all_constr}) as linear combinations of $\phi_J(x)$ with some coefficients $U^J_A$
    \begin{align}
    &\phi_A =\int d^3x \;  U^J_A(x) \, \phi_J(x)=
    \int d^3x\,\big[\,a^j_A(x)P_j(x)\nonumber\\
    &\quad\,\,\,+b^j_A(x)H_j(x)
    +\!c^j_A(x)T_j(x)+d^j_A(x)S_j(x)\,\big].   \label{1st_class_lin}
    \end{align}

Here we use the first part of the alphabet with capital roman letters $A,B,C,...$ to label the first class constraints in contrast to indices from the second part of the alphabet $I,J,...$ labelling the original set of spatially local constraints. The nature of these indices can be very different depending on the choice of basis in the space of $\phi_A$. To begin with, the constraints $\phi_A$ in contrast to $\phi_I$ almost always are spatially nonlocal like, say, irreducible components of transverse vectors and tensors. Therefore, these indices should include both discrete labels with a finite range and continuous or countable labels with an infinite range, like for example a momentum of the Fourier transform $\mathbf k$, $A\mapsto A=1,2,...; \mathbf k$. We will not specify them here, because they might be very different in different cases of closed compact or infinite open space $\varSigma$. We will only assume the DeWitt rule for contraction of these repeated indices, which implies not only the summation over the discrete labels in $A$, but also the integration over its continuous part. 

With the choice of first class constraints in the form (\ref{1st_class_lin}) the coefficients  $U^J_A=a^j_A,b^j_A,c^j_A,d^j_A$ should form on the subspace of vanishing constraints in phase space a zero eigenvalue eigenvector of the matrix of constraint commutators
    \begin{align}
    &\bigl\{ \phi_A, \phi_I(y) \bigr\}\,
    \Big|_{\,\phi_I=0}\nonumber\\
    &\quad\equiv\int d^3x \,
    U^J_A(x)\,\bigl\{ \phi_J(x), \phi_I(y)
    \bigr\}\,\Big|_{\,\phi_I=0}  = 0.              \label{U_equation}
    \end{align} 
Using the following nonvanishing elements of this matrix
    \begin{equation}
    \begin{aligned}
    &\big\{P_i(x), S_j(y)\big\}= - \partial^y_j \,\bigl(\,\varOmega(y) \, \partial_i \delta(y, x) \bigr), \\
    &\big\{H_i(x), H_j(y)\big\}= H_j(x) \, \partial_i \delta(x, y) - (i, x \leftrightarrow j, y), \\
    &\big\{H_i(x), T_j(y)\big\}= - T(y)\,\partial^y_j \big(\, \varOmega(y) \, \partial_i \delta(y, x)\, \big)\\
    &\qquad- \varOmega(y)\, T_j(y) \, \partial_i \delta(y, x)
    - \partial^y_j \bigl( T_i(y) \, \delta(y, x)\, \bigr), \\
    \end{aligned}\nonumber
    \end{equation}

    \begin{equation}
    \begin{aligned}
    &\big\{H_i(x), S_j(y) \big\}= - \partial^y_j\,
    \Big[\,S_i \,\delta(y, x)-\varOmega\partial_i \partial_k N^k\delta(y, x)  \nonumber\\ & \qquad\qquad\quad+\frac{d\varOmega(y)}{d \ln \gamma}\, \Big(\,2 \partial_k N^k - \frac{\pi N}{\sqrt{\gamma}}\Big)(y)\, \partial_i \delta(y, x) \,\Big], \\
    &\big\{T_i(x), T_j(y) \big\}=\partial_i \partial^y_j
    \bigl((wN)(y)\,\big(wN\gamma^{kl}H_k\big)(x)
    \partial_l\delta(x, y)\bigr)\\
    &\qquad\qquad\qquad\quad - (x \leftrightarrow y), \\
    &\big\{T_i(x), S_j(y) \big\}=
    \partial_i \partial^y_j \bigl( \ldots \bigr),
    \end{aligned}
    \end{equation}
(where we do not explicitly specify the last rather complicated commutator because it will not be needed in what follows) one has the set of explicit equations on $U^J_A(x)$
    \begin{equation}
    \begin{aligned}
    & \partial_i (\varOmega \,\partial_j d^j_A) = 0,\\
    & \partial_i \bigl( \varOmega T\partial_j c^j_A \bigr) +\! \int d^3 y\, \{ H_i(x), S(y) \}\,\partial_j d^j_A(y) = 0, \\
    & T \, \partial_i ( \varOmega\partial_j b^j_A ) - \int d^3 y \, \{T_i(x), S(y) \} \, \partial_j d^j_A(y) = 0, \\
    & \partial_i (\varOmega \,\partial_j a^j_A) + \int d^3 y \, \{S_i(x), H_j(y) \} \, b^j_A(y)\nonumber\\
    &\qquad\qquad\quad- \int d^3 y \, \{S_i(x), T(y) \} \, \partial_j c^j_A(y) = 0.
    \end{aligned}
    \end{equation}
The solution of the first three equations gives the condition of transversality of $d^j_A,c^j_A$ and $b^j_A$
    \begin{equation}
    \partial_j d^j_A = 0,\quad \partial_j c^j_A = 0, \quad \partial_j b^j_A = 0,
    \end{equation}
whereas in view of the relation
    \begin{equation}
    \int\! d^3 y \{S_i(x), H_j(y) \} \, b^j_A(y)\! =\! - \partial_i \big[\,\varOmega \,\partial_j (b^j_A \,\partial_k N^k ) - b^j_A S_j \big],
    \end{equation}
the fourth equation takes the form $\partial_i[\,\varOmega \, \partial_j(a^j_A - b^j_A \, \partial_k N^k)\,] = 0$ and has two solutions. The first is $a^j_A = b^j_A \, \partial_k N^k$ when $ b^j_A \neq 0$ and the second is $\partial_j a^j_A = 0$, otherwise. Therefore, the full set of first-class constraints is
    \begin{align}
    &P_A= \int d^3 x \, a^i_A(x)\, P_i(x),    \label{1st_class_def_1}\\
    &H_A= \int d^3 x \, b^i_A(x)
    \bigl(\,H_i(x) + \partial_k N^k(x) \,
    P_i(x)\,\bigr).                           \label{1st_class_def_2}
    \end{align}

Here $a^i_A(x)$ and $b^i_A(x)$ both form a complete set of transversal vectors enumerated by condensed indices which were discussed above. Without loosing generality these two sets can be identified, what we will do in what follows,
    \begin{equation}
    a^i_A(x)=b^i_A(x)=e^i_A(x), \quad \partial_i e^i_A(x) = 0,
    \end{equation}
$e^i_A(x)$ representing some complete basis in the space of all transverse vector fields. In addition to this basis it is useful to introduce the dual basis $e_i^B(x)$ which has the following properties
    \begin{equation}
    \begin{aligned}
    &\int d^3x\,e^i_A(x)\,e_i^B(x)=\delta^B_A,\,\\
    &e^i_A(x)\,e_j^A(y)=\varPi^i_j(x,y),\quad
    \partial_i\varPi^i_j(x,y)=0.  \label{dual_basis}
    \end{aligned}
    \end{equation}
Here $\varPi^i_j(x,y)$ is a projector on the space of transverse vectors. This dual basis  allows one to show that the first class constraints form the subalgebra -- their commutators express as their own linear combinations without contributions of second class constraints. Calculating the commutators of  (\ref{1st_class_def_1})-(\ref{1st_class_def_2}) one can see that in the resulting expressions the original constraints $P_i(x)$ and $H_i(x)$ get functionally contracted (that is with summation over the discrete index $i$ and integration over space) with some transverse vectors $V^i(x)$. Since $\varPi^i_jV^j\equiv\int d^3y\,\varPi^i_j(x,y)\,V^j(y)=V^i(x)$, one can insert into these contractions the projector $\varPi^i_j$ and then use its bilinear decomposition in the basis of $e^i_A$ and its dual. This converts the resulting answer into the linear combinations of $P_A$ and $H_A$ which read
    \begin{align}
    &\{P_A, P_B\} = \{P_A, H_B\} = 0, \label{PB_primary_new}\\
    &\{H_A, H_B\} = H_C\int d^3 x \, e^C_i
    \bigl( e^j_A \partial_j e^i_B - e^j_B
    \partial_j e^i_A \bigr)\nonumber\\
    &\qquad
    + P_C\int d^3 x \, e^C_i
    \partial_j \bigl[\,\partial_k N^k ( e^i_A e^j_B
    - e^i_B e^j_A )\,\bigr].            \label{PB_secondary}
    \end{align}
Similarly we have Poisson brackets of the first-class and second-class constraints
    \begin{align}
    &\{P_A, P_i\}\!=\!\{P_A, H_i\} =
    \{P_A, T_i\}\! =\!\{P_A, S_i\} = 0, \label{PB_primary_P_1} \\
    &\{H_A, P_i\}=
    - \partial_i ( e^k_A P_k), \label{PB_primary_P_2} \\
    &\{H_A, H_i\} =
    - e^k_A \,\partial_k H_i - H_k \,\partial_i e^k_A,\\
    &\{H_A, T_i\} =
    - \partial_i (e^k_A T_k), \\
    &\{H_A, S_i\} = -\partial_i (e^k_A S_k),
    \end{align}
and the Poisson brackets of first-class constraints with the Hamiltonian
    \begin{align}
    &\{\,{\cal H}, P_A\}= \int d^3 x \, e^i_A \, H_i, \label{PB_H_primary}\\
    &\{\,{\cal H}, H_A\}= \int d^3 x\, H_i\,
    \partial_j\bigl[\, N^j e^i_A - N^i e^j_A\,\bigr]\label{PB_H_secondary}
    \end{align}

\section{Canonical realization of Lagrangian symmetries}
According to the conventional canonical formalism of gauge constrained systems \cite{Henneaux-Teitelboim}, the number of Lagrangian gauge symmetries of the theory equals the number of the primary first class constraints, which is the essence of the so-called Dirac conjecture. Under a number of assumptions on the structure of constraint algebra and the distribution of constraints in several generations (primary, secondary, etc.) this conjecture was proven in \cite{Henneaux-Teitelboim}. In our case of GUMG theory in its physically interesting $S_i=0$, $T\neq 0$ branch the situation seems very involved and even contradictory -- the number of anticipated local gauge symmetries of the Lagrangian action is three, while there are four first class constraints. Moreover, these constraints do not belong to a concrete generation, but represent linear combinations of the original primary and secondary constraints. In addition we have second class constraints in all four generations, not to say that the tertiary, $T_i=0$, and quaternary, $S_i=0$, constraints are reducible in view of their longitudinal nature. In this section we will apply a modified version of the Henneaux-Teitelboim procedure \cite{Henneaux-Teitelboim} to recover the Lagrangian symmetries of the theory and show that it actually satisfies the Dirac conjecture, even though it does not satisfy the set of assumptions made in \cite{Henneaux-Teitelboim}. Quite interestingly, this canonical procedure recovers only two gauge symmetries of the GUMG action -- spatial diffeomorphisms with the transverse vector field (\ref{xi_trans}) and prohibits the $\xi^0$-diffeomorphism (\ref{xi_long}) even despite the fact that at the Lagrangian level the latter falls out of the category of local gauge symmetries entirely due to subtle nonlocal behavior at spacetime boundaries.

\subsection{Symmetries of the extended and total actions}
\hspace{\parindent}
The recovery of Lagrangian symmetries of generic constrained systems begins in \cite{Henneaux-Teitelboim} with the construction of the {\em extended} canonical action
    \begin{equation}
    S_E = \int dt \, \big(\,p_k \dot q^k
    - {\cal H} - u^m \phi_m\big),             \label{ext_act_ndef}
    \end{equation}
which includes together with the primary constraints $\phi_{m_1}$ a full set of all the constraints $\phi_m$ and their conjugated Lagrange multipliers $u^m$, consisting of $M$ (primary, secondary, etc.) generations,
    \begin{equation}
    \phi_m = (\,\phi_{m_1}, \phi_{m_2},
    \ldots \phi_{m_M}), \quad m = m_1, \ldots m_M.    \label{generations}
    \end{equation}
Here $m_i$, $i=1,\ldots M$, enumerates the members of the $i$-th generation.
In general, the first class constraints $\phi_A$ do not belong to a concrete generation but rather represent a linear combination of $\phi_m$ mixing the constraints from different generations with some coefficients $U^m_A$ -- zero-vectors of the matrix $\{\,\phi_n, \phi_m \}|_{\,\phi=0}$ (cf. Eq.(\ref{U_equation})),
    \begin{equation}
    \phi_A = U^m_A \phi_m.   \label{U_def}
    \end{equation}
Off the constraint surface in phase space they satisfy the following commutation relations with some structure functions,
    \begin{align}
    \{\, \phi_A, \phi_m \} = C^n_{A\, m} \phi_n, \quad
    \{\, {\cal H}, \phi_A \}= V^m_A \phi_m,
    \end{align}
and allow one to have a canonical transformation of phase-space variables which preserves the constraints and the Hamiltonian
    \begin{equation}
    \delta \begin{pmatrix} \,q^n \\ \,p_n \end{pmatrix}
    = \left\{ \begin{pmatrix} \,q^n \\ \,p_n \end{pmatrix},
    \phi_A\,\mu^A \right\}.                                \label{tr_law}
    \end{equation}
Here $\mu^A$ are some gauge parameters generally depending explicitly on time, phase-space variables and Lagrange multipliers, $\mu^A=\mu^A(t,q,p,u)$.

This canonical transformation can be accompanied by the transformation of the Lagrange multipliers $u^m$ which together with (\ref{tr_law}) leaves the extended action invariant when the parameters $\mu^A$ have in time a compact support. In view of the transformations of the symplectic term 
    \begin{equation}
    \delta\, (\,p_k \dot q^k\,) = \frac{d}{d t} \left[\,\Big( p_k \frac{\partial}{\partial p_k} - 1\,\Big)\,\phi_A \mu^A \,\right] + \phi_A \frac{D \mu^A}{D t},
    \end{equation}
where $D \mu^A/Dt$ denotes the partial time derivative acting only on explicit time dependence on time and Lagrange multipliers
    \begin{eqnarray}
    &&\frac{D\mu^A}{D t} =\Big(\,\partial_t+\dot u^m\frac{\partial}{\partial u^m}\Big)\,\mu^A(t,q,p,u)\nonumber\\
    &&\quad\quad=\Big(\,\frac{d}{d t} - \dot q^k \frac{\partial}{\partial q^k} - \dot p_k \frac{\partial}{\partial p_k}
    \Big)\,\mu^A(t,q,p,u),
    \end{eqnarray}
the variation of the extended action (after omitting the total derivative term) takes the form
    \begin{eqnarray}
    &&\delta S_E = \int dt \, \phi_m \biggl[\, U^m_A \Big(\frac{D \mu^A}{D t} + \bigl\{\mu^A,{\cal H} + u^n \phi_n \bigr\} \Big)\nonumber\\
    &&\qquad\qquad+\,C^m_{A n} \mu^A  u^n - V^m_A \mu^A - \delta u^m  \biggr].
    \end{eqnarray}
This vanishes under an obvious transformation law for the Lagrange multipliers
    \begin{equation}
    \delta u^m  = U^m_A \Big(\frac{D \mu^A}{D t}
    + \bigl\{\mu^A, {\cal H} + u^n \phi_n \bigr\} \Big)
    + C^m_{A n} \mu^A  u^n - V^m_A \mu^A.           \label{tr_law_u}
    \end{equation}
Note that this equation includes as a particular case the situation when the original set of constraints was already splitted into the first and second class constraints. In this case $U^m_A=\delta^m_A$, and the transformations of Lagrange multipliers also split into those of the first class constraints with the explicit time derivative of $\mu^A$ and those of the second class ones without this time derivative.

Transformations (\ref{tr_law}) and (\ref{tr_law_u}) leave the extended action invariant {\em off shell} for arbitrary values of all (primary and higher)  Lagrange multipliers $u^m$, and their number equals the number of {\em all first class constraints}. The number of Lagrangian gauge transformations is smaller and they act on the Lagrangian action functional having a smaller number of arguments -- only Lagrangian coordinates $q^k$. These transformations are in one-to-one correspondence with the transformations of the so-called {\em total} canonical action \cite{Henneaux-Teitelboim} including, in contrast to the extended action (\ref{ext_act_ndef}), only the primary constraints, 
    \begin{equation}
    S_T = \int dt \, (p_k \dot q^k - {\cal H}
    - u^{m_1} \phi_{m_1}).                         \label{tot_action}
    \end{equation}
This action can be regarded as a partial gauge fixing of the gauge symmetries of the extended action in a special gauge of vanishing secondary and higher Lagrange multipliers $u^{m_i} = 0$, $i = 2, \ldots M$. Therefore, its invariance transformations are residual gauge transformations (\ref{tr_law}) and (\ref{tr_law_u}) with the special values of parameters $\mu^A=\mu^A(t,q,p,u)$ preserving this gauge, $\delta u^{m_i} = 0$ for $i = 2, \ldots M$. In view of the above equation for $\delta u^m$ this leads to the relation
    \begin{align}
    &U^{m_i}_A\Big(\frac{D \mu^A}{D t}
    + \bigl\{\mu^A, {\cal H} + u^{m_1} \phi_{m_1} \bigr\} \Big)\nonumber\\ &\qquad+\, C^{m_i}_{A m_1} \mu^A  u^{m_1}
    - V^{m_i} _A \mu^A = 0, \, i = 2, \ldots M, \label{basic_eq}
    \end{align}
which should be treated as a set of equations on $\mu^A$. When this set can be solved for a subset of parameters $\mu^A$ {\em ultralocally in time} in terms of a complementary subset of independent parameters and their time derivatives, then the resulting transformations (\ref{tr_law}) and (\ref{tr_law_u}) with $m=m_1$ will form the gauge symmetries of the total action which are equivalent to the Lagrangian symmetries of the theory. Obviously, the number of these symmetries coincides with the range of this subset of independent parameters $\mu^A$ complementary to the dependent ones.

When all the constraints are of the first class, Eq.(\ref{basic_eq}) can be further simplified. Indeed, in this case $U^m_A$ becomes an identity matrix, so that the indices $m$ and $A$ become identified. Moreover, the procedure of constraints derivation implies the following simplifications of structure functions: $V^{m_i}_{m_j} = 0$ when $i > j + 1$ and $C^{m_i}_{m_j,\,m_1} = 0$ when $i > j$ (following from the fact that the $(i+1)$-th generation of constraint is generated only by the first $i$ generations, and because nonvanishing values of $C^{m_i}_{m_j,\,m_1}$ with $i>j+1$ would imply the equation on the Lagrange multiplier $u^{m_1}$ rather than the new constraint \cite{Henneaux-Teitelboim}). This allows one to rewrite Eq.(\ref{basic_eq}) as the following chain of equations
    \begin{align}
    &V^{m_i}_{m_{i-1}} \mu^{m_{i-1}}
    = \frac{D \mu^{m_i}}{D t}
    + \bigl\{\,\mu^{m_i}, {\cal H}
    + u^{m_1} \phi_{m_1} \bigr\}\nonumber\\
    &\qquad\qquad\qquad+ \sum_{j = i}^M \big(\,C^{m_i}_{m_j m_1} u^{m_1}
    - V^{m_i}_{m_j}\big)\, \mu^{m_j}                   \label{basic_eq_1}
    \end{align}
with $\quad i = 2, \ldots M$, where the summation over generations runs for each $i$ from $i$ to $M$. This system of equations can be solved for $\mu^{m_{i-1}}$ recursively starting from $i = M$ and taking $\mu^{m_M}$ as independent parameters. This procedure ends at the step $i=2$, in which we evaluate $\mu^{m_1}$ in terms of these independent parameters.

\subsection{Symmetries of the GUMG action}
Here we apply the above procedure to two different bifurcating branches of the GUMG theory. The branch of $T=0$ physically is less interesting because it corresponds to the vacuum general relativistic model without the perfect fluid. However, to demonstrate the recovery of spacetime diffeomorphisms restricted by the condition (\ref{condition}), we start with this particular branch and the related case of the unimodular gravity -- both having only first class constraints.

\subsubsection*{\bf Symmetries of $T = 0$ branch}
\hspace{\parindent}
All the constraints $P_i$, $H_i$ and $T$ in the extended action of the $T=0$ branch
    \begin{align}
    &S_E = \int dt \, d^3x \, \bigl(\pi^{ij} \dot \gamma_{ij}
    + P_i \dot N^i - NH_\bot - N^i H_i\nonumber\\
    &\qquad\qquad\qquad\qquad - u_1^i P_i
    - u_2^i H_i - u_3 T \bigr),                     \label{S_E_Tb}
    \end{align}
belong to the first class and, according to the previous section, give rise to extended gauge transformations, which we derive below. For this purpose, we denote $P_i(x)$, $H_i(x)$ and $T(x)$ respectively as $\phi_{m_1 x}$, $\phi_{m_2 x}$ and $\phi_{3x}\equiv\phi_x$ and rewrite the algebra of these constraints (\ref{PB_primary}) and (\ref{PB_H_P})--(\ref{PB_H_T}) in condensed notations with $m_i\mapsto m_1x,\; m_2x,\; m_3x\equiv x$, the index $i=1,2,3$ marking in accordance with (\ref{generations}) the constraints generation. Then in view of (\ref{PB_primary}) $\{ \phi_{m_1 x}, \phi_{m_i y} \}=0$, and the corresponding structure functions turn out to be vanishing, $C^{m_j z}_{m_1 x, \, m_i y} = 0$. In these notations the relations (\ref{PB_H_P})--(\ref{PB_H_T}) read as
    \begin{align}
    &\{ {\cal H}, \phi_{m_1 x} \}
    = V^{m_2 y}_{m_1 x}\, \phi_{m_2 y}, \label{PB_H_P_T=0}\\
    &\{ {\cal H}, \phi_{m_2 x} \}
    = V^{n_2 y}_{m_2 x}\, \phi_{n_2 y}
    + V^{y}_{m_2 x} \phi_{y}, \\
    &\{ {\cal H}, \phi_{x} \}
    =  V^{m_2 y}_{x}\, \phi_{m_2 y}
    + V^{y}_{x}\, \phi_{y},            \label{PB_H_phi_T=0}
    \end{align}
where the integration over $y$ is implied and the remaining structure functions equal
    \begin{equation}
    \begin{aligned}
    &V^{m_2 y}_{m_1 x}= \delta(x, y)
    \delta^{m_2}_{m_1},\\
    &V^{n_2 y}_{m_2 x} =
    - \partial_k \bigl(N^k\delta (x, y)\bigr) \delta^{n_2}_{m_2}
    - \delta(x, y) \, \partial_{m_2} N^{n_2}, \\
    &V^{y}_{m_2 x} = - \partial_{m_2}\delta(x, y), \\
    &V^{m_2 y}_{x}= - w(x) \,
    \partial_k \bigl( N^2 \gamma^{k m_2}
    \delta(x, y) \bigr), \\
    &V^{y}_{x} = - N^k(x) \,
    \partial_k \delta(x, y) - S\,\delta(x, y).
    \end{aligned}\nonumber
    \end{equation}
In the case of solely first class constraints the matrix $U^m_A$ in (\ref{U_def}) is a unit one, $U^m_A=\delta^m_A$, and  Eq.(\ref{basic_eq_1}) for the parameters of residual gauge transformations, $\mu^A\mapsto(\mu_1^{m_1 y},\;\mu_2^{m_2 y},\;\mu_{3}^{y})$, takes the form of the following two equations
    \begin{align}
    &\mu_2^{m_2 y} V^{x}_{m_2 y} = \frac{D \mu_3^x}{Dt}
    + \{\mu_3^x, {\cal H} \} + u^{m_1 y} \{\mu_3^x, \phi_{m_1 y} \}
    - \mu_{3}^{y} V^{x}_{y},                   \label{be1} \\
    &\mu_1^{m_1 y} V^{m_2 x}_{m_1 y}= \frac{D \mu_2^{m_2 x}}{Dt}
    + \{\mu_2^{m_2 x}, {\cal H} \}
    + u^{m_1 y} \{\mu_2^{m_2 x}, \phi_{m_1 y} \}\nonumber\\
    &\qquad\qquad\qquad- \mu_2^{m_2 y} V^{m_2 x}_{m_2 y}
    - \mu_3^{y} V^{m_2 x}_{y}.                     \label{be2}
    \end{align}

When the parameter $\mu_3$ is rewritten in terms of a new independent parameter $\xi^0$, $\mu_3 = \xi^0/w$, the first equation (\ref{be1}) explicitly reads as
    \begin{align}
    &\partial_k \mu_2^k = \mfrac1w \Bigl( \dot \xi^0
    - N^k \partial_k \xi^0 - \bigl[\,\partial_k N^k + \{\ln w, {\cal H}\}\nonumber\\
    &\qquad\qquad\qquad\qquad\qquad - N^k \partial_k \ln w - S\,\bigr]\, \xi^0 \Bigr),
    \end{align}
and in view of the expression for $S$, $S = (1 + w) \partial_k N^k + \{\ln w, {\cal H}\}- N^k \partial_k \ln w$ -- a simple corollary of Eqs.(\ref{S})-(\ref{Omega}), it finally takes the form
    \begin{equation}
    \partial_k (\mu_2^k - \xi^0 N^k)
    = \frac1w \mathcal D_t \xi^0,      \label{1st_eq}
    \end{equation}
where $\mathcal D_t$ is defined in Eq.(\ref{xi_long}), $\mathcal D_t = \partial_t - (1 + w)N^k \partial_k$. One solution of this equation is parameterized by an arbitrary function $\xi^0$,
    \begin{equation}
    \mu_2^i = \xi^0 N^i +
    \partial^i \frac1{w\Delta} \mathcal D_t \xi^0,     \label{1st_sol}
    \end{equation}
while another one is an arbitrary transverse vector with $\xi^0\equiv 0$,
    \begin{equation}
    \mu_2^i = \xi_\bot^i, \quad \partial_k \xi_\bot^k = 0. \label{2nd_sol}
    \end{equation}

The explicit form of the second equation (\ref{be2}),
    \begin{align}
    &\mu_1^i = \frac{D \mu_2^i}{D t} + \Big\{\mu_2^i, \mathcal H 
    + {\textstyle\int} d^3 y \, u_1^k P_k\Big\}
    + \mu_2^k \partial_k N^i\nonumber\\
    &\qquad\qquad\qquad\qquad- N^k \partial_k \mu_2^i - N^2\gamma^{ik}\partial_k (w \mu_3),
    \end{align}
can now be used for the determination of the parameter $\mu_1^i$. For the solution (\ref{1st_sol}) it can be rewritten as
    \begin{align}
    &\mu_1^i= \dot\xi^0 N^i + \xi^0 \, u_1^i
    - \bigl(N^2\gamma^{ij} + N^i N^j\bigr) \partial_j \xi^0\nonumber\\
    &\qquad+\Big\{\partial^i \frac1{w\Delta} \mathcal D_t \xi^0, {\cal H}+ {\textstyle\int}d^3x \, u_1^l P_l \Big\}
    +\partial^i \frac1{w\Delta} \mathcal D_t \dot \xi^0\nonumber\\
    &\qquad+
    \bigl(\partial_jN^i-\delta^i_j N^l\partial_l\bigr)\,
    \partial^j \frac1{w\Delta} \mathcal D_t \xi^0,     \label{1st_sol_P}
    \end{align}
 and
    \begin{equation}
    \mu_1^i = (\partial_t - N^k \partial_k)\,\xi_\bot^i
    +\xi_\bot^k \partial_k N^i                          \label{2nd_sol_P}
    \end{equation}
for the solution (\ref{2nd_sol}).

Thus, we have the following two canonical generators of gauge transformations of the total GUMG action (\ref{action_T}) in the branch $T=0$. The first one is
    \begin{equation}
    G_0 = \int d^3 x \, \Big( P_i \, \mu_1^i
    + H_i \, \mu_2^i + T \, \frac1w \xi^0 \Big), \label{1st_generator}
    \end{equation}
where $\xi^0$ is an independent parameter and the parameters $\mu_1$ and $\mu_2$ are defined in terms of $\xi^0$ respectively by (\ref{1st_sol_P}) and (\ref{1st_sol}). The second one is
    \begin{equation}
    G_\bot = \int d^3 x \, (P_i \, \mu_1^i
    +  H_i \, \xi_\bot^i ),                       \label{2nd_generator}
    \end{equation}
with an arbitrary transverse vector $\xi_\bot^i$ and the vector $\mu_1^i$ defined in terms of $\xi_\bot^i$ by Eq.(\ref{2nd_sol_P}).

To compare the action of these generators on the variables $N^i$ and $\gamma_{ij}$ with their spacetime diffeomorphisms in the Lagrangian formalism, consider first the transformation of $N^i$ by the generator $G_0$, $\delta_0 N^i = \{N^i, G_0\} =\mu_1^i$. Note that for the Lagrangian values of canonical momenta (that is, when the equation of motion for the momenta is satisfied) $u_1^i = \dot N^i$ and the two (fourth and fifth) terms in (\ref{1st_sol_P}) can be rewritten as
    \begin{align}
    &\left\{\partial^i \frac1{w\Delta} \mathcal D_t \xi^0, {\cal H} + {\textstyle\int}d^3x \, u_1^l P_l \right\} +\partial^i\frac1{w\Delta} \mathcal D_t \dot\xi^0\nonumber\\
    &\qquad\qquad\qquad\qquad\qquad\quad=\partial_t \,
    \partial^i \frac1{w\Delta} \mathcal D_t \xi^0,   \label{auxiliary_eq}
    \end{align}
whence
    \begin{align}
    &\delta_0 N^i = N^i \dot\xi^0 + \dot N^i \xi^0 - \bigl(N^2 \gamma^{ij} + N^i N^j\bigr) \partial_j \xi^0 \nonumber\\
    &\qquad\quad+\bigl[\,\delta^i_j \, (\partial_t - N^l\partial_l)
    + \partial_j N^i\,\bigr]\,
    \partial^j \frac1{w\Delta}
    \mathcal D_t \xi^0,                 \label{1st_sol_shift_trans}
    \end{align}
which is just the diffeomorphism transformation law of the shift function (\ref{shift_diffeo}) with the spacetime vector parameter (\ref{xi_long}). A similar calculation for the canonical transformation $\delta_\bot N^i = \{N^i, G_\bot\} = \mu_1^k$ with $\mu_1^k$ given by (\ref{2nd_sol_P})
shows that it is just a diffeomorphism with the parameter (\ref{xi_long}).

Next, considering the gauge transformations of $\gamma_{ij}$ take into account that with the Lagrangian value of the momentum $P_i = 0$ the first term in (\ref{1st_generator}) and (\ref{2nd_generator}) vanishes, so that the commutator of $\gamma_{ij}$ with $\mu_1^i$ does not contribute at all. Moreover, note that the generator of the form $G = {\textstyle\int} d^3x \, (H_\bot F^\bot + H_i \, F^i)$ gives rise to the diffeomorphism of $\gamma_{ij}$ with the vector parameter $\xi^\mu$ defined by the relations $\xi^0 = F^\bot/N$ and $\xi^i = F^i - N^i F^\bot/N$ ($F^\bot$ and $F^i$ are respectively the normal and tangential projections of $\xi^\mu$ on the spatial slice of constant $x^0=t$). Therefore, comparison with the generators (\ref{1st_generator}) and (\ref{2nd_generator}) (remember that $T=wNH_\perp$) shows that $G_0$ defines the diffeomorphism of $\gamma_{ij}$ with the parameter (\ref{xi_long}), while $G_\bot$ defines a spatial diffeomorphism with the parameter (\ref{xi_trans}), which is in a complete agreement with the gauge symmetries of the Lagrangian formalism.

\subsubsection*{\bf Symmetries of the unimodular gravity theory}
The UMG theory, which is directly related to the $T=0$ branch of GUMG, also demonstrates the canonical realization of all three diffeomorphism symmetries. In this case the extended action differs from (\ref{S_E_Tb}) only by the last term which should read as $-u^i_3 T_i$. Similarly, the full set of first class constraints $P_m(x)$, $H_m(x)$ and $T_m(x)$ denoted respectively as $\phi_{m_1 x}$, $\phi_{m_2 x}$ and $\phi_{m_3 x}$ differs from that of the $T=0$ branch by the replacement $\phi_x\to\phi_{m_3x}$. Their algebra again has structure functions with vanishing components $C^{m_j z}_{m_1 x, \, m_i y} = 0$, and their commutators with the Hamiltonian repeat the relations (\ref{PB_H_P_T=0})-(\ref{PB_H_phi_T=0}) with obvious replacements $\phi_x\to\phi_{m_3x}$, $V^y_{m_2 x}\to V^{m_3 y}_{m_2 x}$, $V^{m_2 y}_x\to V^{m_2 y}_{m_3 x}$ and $V^y_x\to V^{n_3 y}_{m_3 x}$, where
    \begin{align}
    V^{m_3 y}_{m_2 x} &= - \delta^{m_3}_{m_2} \, \delta(x, y), \\
    V^{m_2 y}_{m_3 x} &= \partial_{m_3}
    \partial_k \bigl( N^2
    \gamma^{k m_2} \, \delta(x, y) \bigr), \\
    V^{n_3 y}_{m_3 x} &= -\partial_{m_3}
    \bigl( N^{n_3}\, \delta(x, y) \bigr).
    \end{align}
Correspondingly, the equations (\ref{basic_eq_1}) take the form
    \begin{align}
    &\big(\,V^{m_3 x}_{m_2 y}\,\mu_2^{m_2 y}\big)_\|=
    \Big(\frac{D \mu_3^{m_3 x}}{D t}
    + \{\mu_3^{m_3 x}, \mathcal H \}\nonumber\\
    &\qquad\qquad+ \{\mu_3^{m_3 x}, \phi_{m_1 y} \}\,u^{m_1 y}
    - V^{m_3 x}_{n_3 y}\,\mu_3^{n_3 y}\Big)_\| ,   \label{be1_um} \\
    &\mu_1^{m_2 x}
    = \frac{D \mu_2^{m_2 x}}{D t}
    + \{\mu_2^{m_2 x}, \mathcal H \}
    + \{\mu_2^{m_2 x}, \phi_{n_1 y} \}\,u^{n_1 y}\nonumber\\
    &\qquad\qquad\qquad\quad
    - V^{m_2 x}_{n_2 y}\mu_2^{n_2 y}
    - V^{m_2 x}_{n_3 y}\mu_3^{n_3 y},              \label{be2_um}
    \end{align}
where only the longitudinal part of the first vector equation is enforced, since it is contracted in the transformation of the action with the longitudinal vector $T_{m_3}\equiv\partial_{m_3} T$.

Choosing independent longitudinal vector $\mu_3^k$ as $\mu_3^k = \partial^k (1/\Delta) \xi^0$ with an arbitrary scalar parameter $\xi^0$, one finds a general solution of Eq. (\ref{be1_um})
    \begin{equation}
    \mu_2^k = - \partial^k \frac1\Delta \dot \xi^0 + N^k \, \xi^0 + \xi_\bot^k,
    \end{equation}
where $\xi_\bot^k$ is an arbitrary transverse vector. Therefore, it follows from (\ref{be2_um}) that for the $\xi^0$-transformation with $\xi^k_\bot=0$
    \begin{align}
    &\mu_1^i = \dot\xi^0 N^i + \xi^0 \, u_1^i
    - \bigl(N^2\gamma^{ij}
    + N^i N^j\bigr) \partial_j \xi^0\nonumber\\
    &\qquad\quad\quad
    - \bigl[\,\delta^i_j (\partial_t - N^l \partial_l)+ \partial_jN^i\, \bigr]\,
    \partial^j \mfrac1{\Delta} \dot \xi^0,
    \end{align}
where we took into account that in UMG theory $w=-1$, the operator ${\cal D}_t\equiv\partial_t$ is field independent and, therefore, does not contribute to the Poisson bracket commutator term in Eq.(\ref{auxiliary_eq}). Similarly $\mu_1^i = (\partial_t - N^k \partial_k)\,\xi_\bot^i + \partial_kN^i\,\xi_\bot^k$ for transformations with $\xi^0=0$. Thus, as in the $T=0$ branch of GUMG theory, we again obtain two generating functions (\ref{1st_generator}) and (\ref{2nd_generator}) with $T/w=NH_\perp$, which reproduce at the Lagrangian level the diffeomorphisms with volume preserving vector parameters (\ref{xi_long}) and (\ref{xi_trans}), $\partial_\mu\xi^\mu=0$.

\subsubsection*{\bf Symmetries of $T\neq 0$, $S_i = 0$ branch}
Extended action of this GUMG branch is of the form
    \begin{align}
    &S_E = \int dt \, d^3x \, \bigl(\pi^{ij} \dot \gamma_{ij}
    + P_i \dot N^i - N(\gamma) H_\bot - N^i H_i\nonumber\\
    &\qquad\qquad\qquad- u_1^i P_i - u_2^i H_i - u_3^i T_i -u_4^i S_i \bigr),
    \end{align}
where the first-class constraints $\phi_A = (P_{A_1}, H_{A_2})$, $A\mapsto A_1,A_2$, are hidden in the full set of constraints $\phi_{mx} = \bigl(P_{m_1x}, H_{m_2x}, T_{m_3x}, S_{m_4x}\bigr)$ as the linear combinations (\ref{1st_class_def_1}) and (\ref{1st_class_def_2}), $\phi_{A} = U^{m x}_{A} \phi_{m x}$ with the following non-vanishing components of $U^{m x}_{A}$,
    \begin{equation}
    \begin{aligned}
    &U^{m_1 x}_{A_1} = e^{m_1}_{A_1}(x),\quad
    U^{m_1 x}_{A_2} = e^{m_1}_{A_2}(x)\,\partial_k N^k(x),\\
    &U^{m_2 x}_{A_2} = e^{m_2}_{A_2}(x).
    \end{aligned}
    \end{equation}
In view of (\ref{PB_primary_P_1})-(\ref{PB_primary_P_2}) and (\ref{PB_H_primary})-(\ref{PB_H_secondary}) their algebra reads as
    \begin{align}
    &\{P_{A_1}, \phi_{m_1x}\}= \{P_{A_1}, \phi_{m_2x}\}
    = \{P_{A_1}, \phi_{m_3x}\}\nonumber\\
    &\qquad\qquad\qquad\qquad\qquad\qquad=\{P_{A_1}, \phi_{m_4x}\} = 0,              \label{struc_func_0}\\
    &\{H_{A_2}, \phi_{m_1x}\}=
    C^{n_1y}_{A_2,\, m_1x} \phi_{n_1y},          \label{struc_func_1}\\
    &\{{\cal H}, P_{A_1}\}
    = V^{m_2 x}_{A_1} \phi_{m_2 x} ,\\
    &\{{\cal H}, H_{A_2}\}
    = V^{m_1 x}_{A_2} \phi_{m_1 x} + V^{m_2 x}_{A_2} \phi_{m_2 x},
    \end{align}
where
    \begin{align}
    V^{m_2 x}_{A_1} &= e^{m_2}_{A_1}(x),      \label{struc_func_2}\\
    V^{m_2 x}_{A_2} &=
    \partial_k \bigl[\,N^k(x)\, e^{m_2}_{A_2}(x)
    - N^{m_2}(x)\, e^k_{A_2}(x)\, \bigr].       \label{struc_func_3}
    \end{align}
To obtain the generator of canonical transformation which leaves the total action invariant, one should solve in accordance with the Henneaux-Teitelboim procedure the equations (\ref{basic_eq}) for $\mu^A$ in terms of some independent parameters. For $ i = 3, 4$ these equations are satisfied identically since
    \begin{equation}
    U^{m_ix}_A = C^{m_ix}_{A,\, m_1y} = V^{m_ix}_A = 0, \quad i = 3, 4.
    \end{equation}
In the case of $i = 2$ the equation takes the form
    \begin{align}
    &U^{m_2 x}_{A} \left(\frac{D \mu^{A}}{D t}
    + \bigl\{\,\mu^{A}, {\cal H} + u^{m_1 y} \phi_{m_1 y}\, \bigr\} \right)\nonumber\\
    &\qquad\qquad\quad+ C^{m_2x}_{A,\, m_1 y}\, \mu^{A} u^{m_1 y} - V^{m_2 x}_{A} \mu^{A} = 0.
    \end{align}
Using the explicit expressions (\ref{struc_func_2}), (\ref{struc_func_3}) and the fact that $C^{m_2x}_{A,\, m_1 y}=0$ due to Eqs.(\ref{struc_func_0})-(\ref{struc_func_1}), one obtains the relation
    \begin{equation}
    \mu_{1}^m = \dot \xi_\bot^m
    - \partial_k\big(\,N^k \xi_\bot^m - N^m \xi_\bot^k\,\big), \label{param_rel}
    \end{equation}
written down in terms of transverse vectors $\mu_{1}^m (x)$ and $\xi_\bot^m(x)$ which are in one-to-one correspondence with $\mu_1^A$ and $\mu_2^A$
    \begin{equation}
    \begin{aligned}
    &\mu_{1}^m (x)= e^{m}_{A}(x)\,\mu_1^{A},&
    &\xi_\bot^m(x)= e^{m}_{A}(x)\,\mu_2^{A},\\
    &\mu_1^A= \int d^3x \, e^{A}_{m}(x)\,\mu_{\bot 1}^m(x),&
    &\mu_2^A= \int d^3x \, e^{A}_{m}(x) \, \xi_\bot^m(x).
    \end{aligned} \nonumber
    \end{equation}

Therefore, the canonical generator $G_\bot = P_A \, \mu_1^A + H_A \, \mu_2^A$, with $\mu_1^A$ and $\mu_2^A$ expressed via (\ref{param_rel}) in terms of an arbitrary transverse vector $\xi^i_\perp$ can be rewritten in the reducible basis of transverse vectors as
    \begin{equation}
    G_\bot = \int d^3x \, \big[\,(\dot \xi_\bot^i
    - N^k \partial_k \xi_\bot^i
    + \xi_\bot^k \partial_k N^i )\,P_i
    + \xi_\bot^i H_i\, \big].                \label{2nd_gen_S}
    \end{equation}
Obviously this is just the generator (\ref{2nd_generator}) with the vector $\mu_1^i$ given by (\ref{2nd_sol_P}), derived above for the case of $T=0$ branch, which is responsible for transverse spatial diffeomorphisms of both $\gamma_{ij}$ and $N^i$.

But in contrast to the $T=0$ branch, this is the only set of two (per space point) local gauge transformations that exist in the $S_i=0$, $T\neq 0$ branch of the theory. The number of local gauge transformations coincides with the number of {\em primary} first class constraints, which is less than four -- the full number of first class constraints. The Dirac conjecture turns out to be true again, even though the set of assumptions listed in \cite{Henneaux-Teitelboim} as sufficient conditions for the validity of this conjecture are not satisfied. As one can see, this conclusion based on the canonical formalism of the model fully matches with the Lagrangian formalism accounting for the subtleties of gauge transformations at timelike and spacelike boundaries of spacetime, considered above in Sect.2. Below we reveal the detailed mechanism of this gauge invariance (and its violation) in the approximation of linearized theory on spatially closed and asymptotically flat Friedmann background. 

\section{Linearized theory}
We consider now the GUMG model in the linearized approximation on the background of the homogeneous Friedmann metric of positive or zero spatial curvature, $k=+1$ or $k=0$ respectively. In these cases we have the metric and curvature in terms of the scale factor $a(t)$ and correspondingly the metric $\sigma_{ij}$ of the 3-dimensional sphere of unit radius or the flat metric,
    \begin{eqnarray}
    &&\gamma_{ij}=a^2(t)\sigma_{ij}(x),\quad N=N(a),
    \quad N^i=0,\quad ,                          \label{Friedmann_back1}\\
    &&{}^3 R=\frac6{a^2}\,k,\quad
    K_{ij}=-\frac{a\dot a}N\,\sigma_{ij},
    \quad H=\frac{\dot a}a.                        \label{Friedmann_back2}
    \end{eqnarray}
$H$ denotes the Hubble factor of the Friedmann background -- we hope that it will not be confused with the notation for the GR Hamiltonian and momentum constraints. Equations of motion for this background read
    \begin{align}
    &\frac{\delta S}{\delta \gamma_{ij}}
    =\sqrt{\sigma} \frac{a^3}{N}
    \Big[\, 2 \dot H + 3 (1- w) H^2\nonumber\\
    &\qquad\qquad\qquad+ (1 + 3 w) \frac{N^2}{a^2}k\, \Big] \sigma^{ij}=0,  \label{hom_eq}\\
    &\frac{\delta S}{\delta N^i}\equiv 0,
    \end{align}
where in view of homogeneity the shift component is identically satisfied. One can check that the first equation has the integral of motion with a constant $C$
    \begin{eqnarray}
    \frac{H^2}{N^2}+\frac{k}{a^2}=\frac{C}{3Na^3},    \label{C}
    \end{eqnarray}
which can be interpreted as the Friedmann equation with the physical Hubble factor in cosmic time, $H/N=\dot a/Na$, and the dark energy density $\varepsilon=2C/Na^3$ (the right hand side being $8\pi G\varepsilon/3$ in our units with $G=1/16\pi$). The nonzero constant $C$ is what distinguishes the $T\neq 0$ branch of the model from its general relativistic branch and simulates this dark energy density which identically satisfies the stress tensor conservation law
    \begin{eqnarray}
    \frac{d\varepsilon}{da}=
    -3\,(1+w)\,\frac\varepsilon{a}.        \label{epsilon_equation}
    \end{eqnarray}

\subsection{Action of linearized theory and its gauge invariance properties}
Now we expand the GUMG action to second order in perturbations of $\gamma_{ij}$ and $N^i$, which we denote as follows
    \begin{align}
    \delta \gamma_{ij} =  a^2 s_{ij}, \quad
    \delta N^i =  S^i,  \label{perturbations}
    \end{align}
along with the first and second order variations of $N(\gamma)$ -- the function of the perturbed 3-metric,
    \begin{align}
    \delta N = N A,\quad \delta^2 N= \frac14 w N
    \big[\, (\varOmega - 1) s^2 - 2s_{ij}^2 \big],   \label{perturbations1}
    \end{align}
Here in view of $\delta N=wNs/2$
    \begin{equation}
    A=\frac12\,w\,s,\quad s=\sigma^{ij}s_{ij},\quad
    s_{ij}^2\equiv\sigma^{im}\sigma^{jn}s_{ij}s_{mn},
    \end{equation}
$\varOmega$ is defined by Eq.(\ref{Omega}), and everywhere here and in what follows all spatial indices are raised and lowered by the metric $\sigma_{ij}$. As a result the second order variation of the action reads
    \begin{align}
    &\frac12 \delta^2 S_{GUMG} = S_{(2)}\nonumber\\
    &\qquad\qquad+\int dt\,d^3x\, \frac{d}{dt}\Bigl\{\sqrt{\sigma}\frac{a^3H}{N}
    \big(s_{ij}-\frac12 s^2\big)\Bigr\},             \label{ADM_quadr}
    \end{align}
with the first term given by
\begin{widetext}
    \begin{eqnarray}
    &&S_{(2)} = \int dt \, d^3 x\, \sqrt{\sigma} \,a^3\, \left\{\,\frac1N\Bigl(\nabla_{(i} S_{j)} -\frac{\dot s_{ij}}2 \Bigr)^2
    -\frac1N\Bigl(\nabla_k S^k - \frac{\dot s}2 \Bigr)^2 - \frac{4H}N \Bigl(\nabla_k S^k - \frac{\dot s}2 \Bigr) A\right. \nonumber\\
    &&\qquad\qquad\quad
    -\frac6N\, H^2 A^2+\frac14\, \frac{N}{a^2} \big(\,2 \nabla^k s^{ij} \nabla_i s_{jk} - \nabla^k s^{ij} \nabla_k s_{ij} - 2 \nabla_i s^{ij} \nabla_j s + \nabla^i s \nabla_i s\,\big)\nonumber\\
    &&\qquad\qquad\quad+\frac{N}{a^2} k\, \Bigl(s_{ij}^2-\frac12 s^2\Bigr)-2\frac{N}{a^2} k s A-\frac{N}{a^2} (\nabla^j s_{ij} - \nabla^i s) \nabla_i A \left.
    +\frac1{4a^3}\,Cw\varOmega\,s^2\right\},              \label{S_quadr}
    \end{eqnarray}
\end{widetext}
where in addition to the rule of raising and lowering indices by the metric $\sigma_{ij}$ all covariant derivatives are also defined with respect to this metric. One can check that all the terms except the last one coincide with the quadratic part of the Einstein action on the Friedmann background \cite{Garriga_etal}. The last term is a modification due to GUMG generalization, which obviously vanishes for $C=0$ (GR branch of the model) and for $w=-1$ when $\varOmega=0$ -- see Eq.(\ref{Omega}). In this case this is just a unimodular gravity model.

Below we will discard the total derivative term in (\ref{ADM_quadr}) and build the canonical formalism and the physical sector for the quadratic action $S_{(2)}$. Hamiltonian formalisms for (\ref{ADM_quadr}) and (\ref{S_quadr}) differ, of course, by a canonical transformation, but for $S_{(2)}$ the formalism is simpler. Another remark concerns spatial surface terms in (\ref{S_quadr}). The spatial derivatives here are organized so that the variational principle is consistent under a fixed metric of the timelike boundary (only squares of first order derivatives and no second order derivatives are contained in $S_{(2)}$). On the other hand, for spatially closed model ($k=1$) this problem is irrelevant and the same applies to asymptotically flat space, because falloff condition for metric perturbations makes all quadratic surface terms arising from integrations by parts vanishing at infinity.

As usual in the linearized theory on the field background, the gauge transformations of perturbation fields are defined as the part of the transformations (\ref{lapse_diffeo})-(\ref{gamma_diffeo}) of zeroth order in perturbations (their cross terms with $\xi^\mu$ belong to the next order of perturbation theory). Therefore, these transformations read
    \begin{align}
    &\delta^\xi s_{ij}= 2 H \sigma_{ij}\,\xi^0
    + \nabla_i \xi_j+ \nabla_j \xi_i,\quad \xi_i=\sigma_{ij}\xi^j, \\
    &\delta^\xi S^k= - \frac{N^2}{a^2} \nabla^k \xi^0 + \dot\xi^k,\quad
    \nabla^i=\sigma^{ij}\nabla_j,
    \end{align}
and they of course reproduce for the vectors $\xi^\mu$ satisfying the ($\sigma$-metric covariantized) equation (\ref{xi_equation}) the general relativistic transformation law for $A=ws/2$, $\delta^\xi A= \partial_t(N \xi^0)/N$. Under these transformations the quadratic action transforms as
\begin{widetext}
    \begin{eqnarray}
    &&\delta^\xi S_{(2)} = \int dt  \int_{\partial\varSigma} d^2 \varSigma_i \,\sqrt{\sigma} \, Na \, \Big[\, 2\, \bigl(\,S^i (\Delta + 2k) - S_j \nabla^i \nabla^j \, \bigr)\, \xi^0+ (\dot s^{ij}\nabla_j
    - \dot s \, \nabla^i)\, \xi^0\nonumber\\
    &&\qquad\qquad + 4 H \, A\, \nabla^i\xi^0 +s^{jk} \nabla^i\nabla_j \xi_k
    -s^{ij} \Delta\xi_j-4 k\,A\,\xi^i
    -\frac{2C}{Na}\, (1 + w)\, S^i \, \xi^0
    \, \Big]\nonumber\\
    &&\qquad\qquad+\int d^3 x\, \sqrt{\sigma}\, N a \Big[\,\bigl(- \nabla_j s^{ij} \, \nabla_i + \nabla^i  s
    \,\nabla_i  - 2 k s \bigr) \, \xi^0+\frac{C}{Na}\, (1 + w)\, s \, \xi^0\, \Big]\,\Big|_{\,t_-}^{\,t_+}\nonumber\\
    &&\qquad\qquad+\,2\,C \int dt \, d^3 x\,\sqrt{\sigma}\, \Big[\, \frac{d \ln w}{d \ln \gamma}\, s \,
    \dot \xi^0 + (1 + w)\, S^i\, \nabla_i \xi^0\,\Big],   \label{act_var}
    \end{eqnarray}
\end{widetext}
where we carefully took into account all surface terms originating from integration by parts.

As is expected this expression for general relativistic case of $C=0$ reduces to the surface integral and vanishes for $\xi^\mu$ with a compact support inside the spacetime domain, which is the case of the theory invariant under local gauge transformations. For GUMG theory we have additional terms including spacetime volume integrals. This is strange because the gauge transformations of GUMG theory are just a subclass of gauge transformations of general relativity, so that the difference in the gauge transformation of the action can only be of the boundary terms type. It turns out indeed that this extra spacetime integral reduces to the surface integral at timelike and spacelike boundaries.

To see this perturb the exact equation (\ref{condition}) for $\xi^\mu$ and obtain the relation
    \begin{align}
    &w\,\nabla_i\xi^i_{(1)}=-\Big[\,\frac{d \ln w}{d \ln \gamma}\, s \, \dot \xi^0_{(0)}\nonumber\\
    &\qquad\qquad\qquad+ (1 + w)\, S^i\,
    \nabla_i \xi^0_{(0)}\,\Big]
    +\dot\xi^0_{(1)}              \label{xi_equation2}
    \end{align}
where $\xi^i_{\scriptscriptstyle(1)}=\delta\xi^i$ and $\xi^0_{\scriptscriptstyle(1)}=\delta\xi^0$ are the first order perturbations of $\xi^\mu$ under the variations of the metric and shift functions on the Friedmann background (\ref{Friedmann_back1})-(\ref{Friedmann_back2}), and we took into account that the background value of $w$ is independent of space coordinates. Therefore, the last spacetime integral term in (\ref{act_var}) also becomes a surface term
    \begin{align}
    &2C\! \int dt \, d^3 x\,\sqrt{\sigma}\, \Big[\, \frac{d \ln w}{d \ln \gamma}\, s \,
    \dot \xi^0 + (1 + w)\, S^i\, \nabla_i \xi^0\,\Big]\nonumber\\
    &\,\,=
    2C\!\int\limits_{\varSigma} d^3x\,\sqrt\sigma\,\xi^0_{(1)}\,\Big|_{\,t_-}^{\,t_+}\!
    -2C\!\int dt\,w\int\limits_{\partial\varSigma}
    d^2\varSigma_i\,\xi^i_{(1)}.                   \label{xi_equation3}
    \end{align}
As we show now, this relation implies that for both closed compact and open asymptotically Friedmann models the action is not invariant under $\xi^0$-transformation.

Indeed, for closed models timelike surface integrals are absent, but $\xi^0$ cannot be a field independent variable with a compact support, because it should satisfy the equation (\ref{xi_equation}) nontrivially depending on the fields. When expanded in powers of metric perturbations, $\xi^\mu=\sum_n\xi^\mu_{\scriptscriptstyle(n)}$, this equation implies that the spatially constant mode of the lowest order $\xi^0_{\scriptscriptstyle(0)}$ is time independent, $\int d^3x\sqrt\sigma\dot\xi^0_{\scriptscriptstyle(0)}=0$, other harmonics of $\xi^0_{\scriptscriptstyle(0)}$ being arbitrary and admitting a compact support in time. On the contrary, the same mode of $\xi^0_{\scriptscriptstyle(1)}$ is strongly restricted by the differential in time equation
    \begin{align}
    &\int d^3 x \sqrt{\sigma} \, \Big[\, \dot \xi^0_{(1)}
    - \frac{d \ln w}{d \ln \gamma} s\, \dot \xi^0_{(0)}\nonumber\\
    &\qquad\qquad\qquad
    - (1+w) S^k \nabla_k  \xi^0_{(0)}\Big] = 0,     \label{par_dep_closed}
    \end{align}
so that $\xi^0_{\scriptscriptstyle(1)}$ cannot have a compact support. In particular,  $\xi^0_{\scriptscriptstyle(1)}(t_\pm)\neq 0$ and the gauge transformation (\ref{act_var}) is nonzero in view of the first term of Eq.(\ref{xi_equation3}). Other terms are vanishing because they already contain one power of perturbation fields, so that $\xi^0$ factor should include only the leading order part $\xi^0_{\scriptscriptstyle(0)}$ which has a compact support in time. Thus, the closed model action is invariant only under the transverse spatial diffeomorphisms with the vectors $\xi^\mu=(0,\xi^i_\perp)$ having a compact support in time. These are the only gauge transformations generated by the canonical first class constraints in the $T\sim C\neq 0$ branch of the theory.

In the open models case with asymptotically flat space slices, $\xi^0$ can be taken field independent, because the integral of equation (\ref{xi_equation2}) is no longer a restriction on the choice of $\xi^0$, but just the relation which determines the flux of the vector $\xi^i_{\scriptscriptstyle(1)}$ through the remote spatial boundary, $\int_\infty d^2\varSigma_i\,\xi^i_{\scriptscriptstyle(1)}$. Therefore, $\xi^0$ can have a compact support in time, and all spacelike integrals at $t_\pm$ in (\ref{act_var}) and (\ref{xi_equation3}) vanish. However, $\xi^i_{\scriptscriptstyle(1)}(x)\sim1/|x|^2$, $|x|\to\infty$, and this ``side" surface flux of the form (\ref{asymp_int}), generated by the last term of (\ref{xi_equation3}), is nonvanishing. Note that this is also the only term that breaks gauge invariance of the action, because other ``side" surface integrals are zero in view of the falloff conditions for $s_{ij}(x),S^i(x),\xi^\mu(x)$.\footnote{Integrability condition for terms without spatial derivatives of metric perturbations in (\ref{S_quadr}) imply that $s_{ij}(x)\sim 1/|\,x\,|^2$, $S^i(x)\sim 1/|\,x\,|$, $\xi^i(x)\sim1/|\,x\,|$ and $\xi^0(x)\sim1/|\,x\,|^2$ (the latter two restrictions guarantee that the gauge transformation (\ref{gamma_diffeo}) with $\dot\gamma_{ij}\sim \delta_{ij}\dot a/a$ should not violate the falloff condition for $s_{ij}(x)$), which guarantees that all surface terms except the one in (\ref{xi_equation3}) vanish at spatial infinity $|\,x\,|\to\infty$.}

Thus, in both closed and open cases the $\xi^0$-transformation of the action is given by the nonvanishing 3-dimensional integral over the full boundary of spacetime (\ref{xi_equation3}),
    \begin{equation}
    \begin{aligned}
    &\delta^\xi S_{(2)}=2\,C\int\limits_{\partial ({}^4\! M)}d^3\varSigma_\mu\,\varXi^\mu_{(1)}, \\
    &{}^4\!M=[\,t_-,t_+]\times {}^3\!\varSigma,
    \quad \varXi^\mu_{(1)}
    =\big(\,\xi^0_{(1)},-w\,\xi^i_{(1)}\big),          \label{act_var1}
    \end{aligned}
    \end{equation}
with a zero timelike or ``side" part in a closed model case and a zero spacelike part, $\varXi^0_{(1)}=0$, for an asymptotically flat model. The GUMG action is gauge invariant only under two spatial diffeomorphisms with a transverse 3-vector having a compact support both in space and time. All this is in full accordance with the canonical formalism of the $T\sim C\neq 0$ branch, considered above. Note that UMG theory with $w=-1$ is an exception from this rule even for $C\neq 0$, because the equation (\ref{condition}) in this case is field independent, and $\xi^\mu_{\scriptscriptstyle(1)}=0$.

\subsection{Physical sector}
The number of degrees of freedom in the physical sector is determined by the number of first and second class constraints and, therefore, is different in different branches of the model. In the $T\neq 0$ branch their counting gives $18-2\times 4- 4=2\times 3$ -- three local degrees of freedom, where 18 is the number of phase space variables $N^i,P_i,\gamma_{ij},\pi^{ij}$, 4 is the number of first class constraints $P_A,H_A$ and 4 is the number of second class constraints -- two longitudinal components of $P_i$ and $H_i$ and two independent constraints among $T_i$ and $S_i$ (since $T_i=\partial_iT$ and $S_i=\partial_iS$ only one longitudinal component counts for each of them). In the $T=0$ branch of GUMG and in UMG, just like in GR, the counting is different and obviously gives two degrees of freedom, $18-2\times 7=2\times 2$, where 7 is the full set of first class constraints $P_i,H_i$ and $T$ \cite{UMG2}.

In the linearized theory disentangling these degrees of freedom runs via the decomposition of the metric perturbations into the linear combination of transverse-traceless tensor modes $t_{ij}$, transverse vector modes $F_i,V_i$ and the scalar modes $\psi, E, B$,
    \begin{align}
    &s_{ij}=t_{ij}+2\nabla_{(i} F_{j)}
    -2 \psi\,\sigma_{ij}+ 2 \nabla_i \nabla_j E,
      \label{lin_metr}\\
    &t^i{}_i = \nabla^i t_{ij}=\nabla_i F^i = 0,\\
    &S_i= V_i+\nabla_i B, \quad \nabla_i V^i = 0.
    \end{align}
Using this decomposition in (\ref{S_quadr}) one finds that the tensor transverse-traceless modes decouple from the vector and scalar sectors,
$S_{(2)}=S_t+S_v+S_s$, in the form of the action of two field-theoretical oscillators on the non-static Friedmann background
    \begin{align}
    \!\!\!\!\!S_t =\!\int dt\, d^3 x \sqrt{\sigma} a^3
    \left[\frac{\dot t_{ij}^2}{4N}
    -\frac{N}{4a^2}(\nabla_kt_{ij})^2
    -\frac{kN}{2a^2}\,t_{ij}^2\right].
    \end{align}
Transverse vector modes $F_i$ and $V_i$ enter only the action $S_v$, while the action of the scalar sector $S_s$ equals
\begin{widetext}
    \begin{eqnarray}
    &&\nonumber\\
    &&S_s =\int dt \, d^3 x \sqrt{\sigma} a^3
    \left[\,-\frac6N (\dot \psi + H A)^2
    -\frac4N\,(\dot \psi + H A)\Delta(B - \dot E)
    + \frac{2k}N (B - \dot E) \Delta (B - \dot E)\right.\nonumber\\
    &&\qquad\qquad\quad\left.- \frac{2N}{a^2}\,
    \psi\,(\Delta + 3 k)\psi+\frac{4N}{a^2}\,A \,(\Delta+3k)\psi\,\right]
    + C \int dt \, d^3 x \sqrt{\sigma}
    \frac{\varOmega}{w} \, A^2,                 \label{scalar_action1}
    \end{eqnarray}
\end{widetext}
where the perturbation of the lapse function $A$, cf. Eq.(\ref{perturbations}), reads in terms of the scalar modes of the metric perturbations as
    \begin{equation}
    A = w (\Delta E - 3 \psi).
    \end{equation}

Obviously, it is this sector that contains in the case of the $T\neq 0$ GUMG branch the third physical degree of freedom. To disentangle it one has to build the canonical formalism of the linearized theory and solve the relevant constraints and canonical gauge conditions in a conventional Hamiltonian reduction procedure. Four gauge conditions, necessary to gauge out the symmetries generated by the four first class constraints $P_A$ and $H_A$ at the linearized level are used for the determination of the four transverse vector modes $(F_i,V_i)$. But their quadratic action $S_v$ vanishes on the linearized constraints, and their choice dynamically affect neither the transverse-traceless, nor the scalar sector of the theory. Therefore, we do not specify vector modes at all and consider only the scalar sector in what follows. The reduction of this scalar sector to physical modes follows only from solving its canonical first and second class constraints -- the linearized version of the constraints of the full nonlinear theory considered above.

As one can see now, this reduction goes very differently in the spatially homogeneous and inhomogeneous subspaces of the model. To begin with, for spatially constant modes $\psi_0$, $E_0$ and $B_0$ the quadratic action (\ref{scalar_action1}) reads as
    \begin{eqnarray}
    &&S_s^{(0)} = \int dt \,d^3x\,\sqrt\sigma\, a^3\, \left\{ - \frac6N\,\big(\,\dot \psi_0-3 w H \psi_0\big)^2\right.\nonumber\\
    &&\qquad\left.+ \Big[\,\frac{9 C}{a^3}\, w\,\Omega
    -\frac{6N}{a^2}k (1 + 6 w)\,\Big] \,
    \big(\psi_0\big)^2 \right\},   \label{GUMG_quad_comp}
    \end{eqnarray}
and does not generate any constraints. This is the action of one global physical mode which is in fact the first order perturbation of the background solution of Eq.(\ref{hom_eq}). This can be easily verified by noting that $S_s^{(0)}$ coincides with the second order variation of the minisuperspace action
    \begin{equation}
    S_s^{(0)}=\frac12\,\delta^2\int dt \,d^3x\,
    \sqrt\sigma\,  \Bigl[\,6\,k\, N(a)\, a
    - \frac{6\,a^3}{N(a)} \frac{\dot a^2}{a^2}\, \Bigr]
    \end{equation}
under the variations of the scale factor $\delta a=a\psi_0$ and the $a$-dependent lapse function $\delta N(a)=3wN(a)\psi_0$. This action generates the second order equation of motion for $\psi_0$ which has a runaway solution corresponding to either the cosmological acceleration or inflation, if it is applied in context of inflationary cosmology. This single mechanical (rather than field-theoretical) mode has a ghost nature because of the negative sign of its kinetic term and does not differ much from the scale factor mode in GR, except that $\psi_0$ in GUMG is dynamically independent and its nonvanishing constant of motion $C$ is freely specified by initial conditions.

For spatially inhomogeneous modes with $\Delta\neq 0$ the situation is different -- this is the case of constrained system.  The definition of the canonical momenta,
    \begin{align}
    \varPi_\psi &= - 4 \sqrt{\sigma}\, \frac{a^3}N \,\big[ \,3 (\dot \psi + H A) + \Delta (B - \dot E)\, \big], \\
    \varPi_E &= 4 \sqrt{\sigma}\, \frac{a^3}N \,
    \big[\, \Delta (\dot \psi + H A)
    - k\, \Delta (B - \dot E)\, \bigr],\\
    \varPi_B&=0,                               \label{Pi_B_zero}
    \end{align}
implies one primary constraint. On the subspace of invertible operator $\Delta+3k$, where the equations
    \begin{align}
    (\Delta + 3k) (\dot \psi + H A) &= \frac N{4a^3}
    \frac1{\sqrt{\sigma}} (\varPi_E - k \varPi_\psi),       \label{pi_s1}\\
    (\Delta + 3k) \,\Delta (B - \dot E) &=
    -\frac N{4a^3} \frac1{\sqrt{\sigma}}
    (\Delta \varPi_\psi + 3 \varPi_E)                    \label{pi_s2}
    \end{align}
can be uniquely solved for $\dot\psi$ and $\dot E$, the Legendre transform with respect to these velocities gives the total canonical action which includes the Hamiltonian $\mathcal H$ and this primary constraint with the Lagrange multiplier $u$,
\begin{widetext}
    \begin{eqnarray}
    &&\nonumber\\
    &&S_s^{(>0)} = \int dt \,\left\{\int d^3x\,
    \big(\,\varPi_\psi \dot \psi + \varPi_E \dot E
    + \varPi_B \dot B\,\big) - \mathcal H
    - \int d^3x\,u\,\varPi_B \right\},             \label{scalar_action2}\\
    &&\mathcal H = \frac N{8a^3}
    \int \frac{d^3x }{\sqrt{\sigma}}
    \Bigl\{-\varPi_\psi \frac{k}{\Delta+3k}\varPi_\psi
    + 2\,\varPi_\psi \frac1{\Delta+3k} \varPi_E
    +  3\,\varPi_E \frac1{\Delta+3k}
    \frac1{\Delta}\varPi_E\, \Bigr\}    \nonumber\\
    &&\qquad\quad+ \int d^3x\,
    \Bigl\{\,2\, N\,a\, \sqrt\sigma\,\psi (\Delta+3k) \psi
    - A\, \Big[\, H \varPi_\psi
    + 4\,\sqrt\sigma\, N\, a\,(\Delta+3k)\,\psi\, \Big]\nonumber\\
    &&\qquad\qquad- C\,\sqrt\sigma\,\frac{\varOmega}{w}\, A^2+ B\,\varPi_E\,\Bigr\}.               \label{Hamiltonian_lin}
    \end{eqnarray}
\end{widetext}

The conservation of the primary constraint (\ref{Pi_B_zero}) leads to the sequence of secondary, tertiary and quaternary constraints
    \begin{eqnarray}
    &&\{\,\varPi_B, \mathcal H\, \} = -\varPi_E = 0, \label{secondary}\\
    &&\{ \,\varPi_E,\mathcal H\, \} = T=0,\\
    &&T\equiv\sqrt\sigma\Delta \Bigl(\,2C\varOmega\, A\nonumber\\
    &&\qquad\quad\quad+ w\,\Bigl[\,H \frac{\varPi_\psi}{\sqrt\sigma}
    + 4Na(\Delta+3k)\psi\,\Bigr] \Bigr),\label{tert1}\\
    &&\frac{\partial T}{\partial t}
    +\{\,T,\mathcal H\,\}=Cw\sqrt{\sigma}\Delta\Bigl(12 H \frac{d \varOmega}{d \ln \gamma}
    \frac{A}w \nonumber\\
    &&\qquad\quad\quad+ \frac{N}{a^3}\,
    \frac{d \ln w}{d \ln \gamma}\,\frac{\varPi_\psi}{\sqrt{\sigma}}
    + 2 \varOmega\,\Delta B \Bigr) = 0,          \label{quat1}
    \end{eqnarray}
where the explicit time derivative $\partial/\partial t$ acts only on the background variables in $T$. The conservation of the last constraint leads to the equation on the Lagrange multiplier $u$, so that the sequence of these constraints terminates at the quaternary one.\footnote{These constraints are equivalent to but not directly coincide with the longitudinal parts of linearized constraints $H_i$, $T_i$, and $S_i$ of the full nonlinear theory considered above. This is because the canonical formalisms of the full theory and its linearized version are related by the canonical transformation corresponding to the total derivative term in (\ref{ADM_quadr}) -- see discussion following Eq.(\ref{S_quadr}).}

There is an additional subtlety for a spatially closed model, $k=+1$, when the operator $\Delta+3$ has discrete zero modes, and the nonlocal terms in the Hamiltonian (\ref{Hamiltonian_lin}) seem to be ill defined. This is the set of four eigenmodes of the Laplacian operator on the 3-sphere, $\Delta Z_{n}(x)=-n(n+2)Z_{n}(x)$, labeled by $n=1$, which have the property $\nabla_i\nabla_jZ_1(x)=-\sigma_{ij}Z_{1}(x)$. This mode, however, does not contribute to the physical sector of the theory. To see this, note that in the subspace of this mode the left hand sides of Eqs.(\ref{pi_s1})-(\ref{pi_s2}) are vanishing, which means that we have extra primary first class constraint
    \begin{equation}
    \varPi_E^{1}-\varPi_\psi^{1}=0.
    \end{equation}
This constraint generates the gauge transformation $E_{1}\to E_{1}+\zeta$, $\psi_{1}\to\psi_{1}-\zeta$, which leaves the linearized 3-metric (\ref{lin_metr}) invariant because its scalar part equals $-2 \psi_{1}\,\sigma_{ij}+2\nabla_i\nabla_jE_{1} =
-2\sigma_{ij}(\psi+E)_{1}$. Indistinguishability of these two terms in the decomposition of $s_{ij}$ into irreducible components requires additional gauge condition to fix $\psi_{1}$ and $E_{1}$ separately, while the physical sector depends only on their sum. The dynamical effect of this extra constraint is that the kinetic term of the $n=1$ mode completely vanishes in the Hamiltonian, so that ill-defined nonlocal terms disappear. Moreover, in view of the secondary constraint $\varPi_E^{1}=0$ the both momenta in the $n=1$ sector vanish, and the tertiary and quaternary constraints completely kill its physical mode, because both $(\psi+E)_{1}$ and $B_{1}$ turn out to be zero due to (\ref{tert1}) and (\ref{quat1}).\footnote{For the open model with $k=0$ the difficulty of the above type becomes the problem of the dipole mode which seems to invalidate the Hamiltonian (\ref{Hamiltonian_lin}) because of the infrared divergent integral $\int d^3x\,\varPi_E(1/\Delta^2)\varPi_E=\int d^3x\,[(1/\Delta)\varPi_E]^2$ with $(1/\Delta)\varPi_E(x)\sim 1/|x|$ at $|x|\to\infty$. But again in the physical sector, when the secondary constraint $\varPi_E=0$ is enforced, this problem is easily circumvented.}

Thus, for spatially inhomogeneous modes with $\Delta\neq 0$ the tertiary and quaternary constraints (\ref{tert1})-(\ref{quat1}) can be algebraically solved for $A$ and $B$ in terms of $\psi$ and $\varPi_\psi$ and used in (\ref{scalar_action2}) along with primary and secondary constraints. This gives the sector of one local scalar degree of freedom with the physical Hamiltonian $\mathcal H_*[\,\psi, \varPi_\psi]$
    \begin{align}
    &\!\!\!\!\!S_s^{(>0)}[\,\psi, \varPi_\psi] =
    \!\int dt\left\{\int d^3 x\,\varPi_\psi \dot \psi
    - \mathcal H_*[\,\psi, \varPi_\psi] \right\},\\
    &\!\!\!\!\!\mathcal H_*[\,\psi, \varPi_\psi] =\int d^3 x\,\Big[\,\frac12\,\varPi_\psi G(\Delta) \varPi_\psi+
    \varPi_\psi V(\Delta) \psi\nonumber\\
    &\qquad\qquad\qquad\qquad\qquad\qquad\quad
    +\frac12\,\psi\,U(\Delta) \psi\,\Big],        \label{H_phys}
    \end{align}
where the coefficienss $G(\Delta)$, $U(\Delta)$ and $V(\Delta)$ are the following (non)local operators -- functions of the covariant Laplacian\footnote{Local degrees of freedom with a nonlocal Hamiltonian sounds as an euphemism, but this is a usual terminology accepted to distinguish field-theoretical modes associated with location in coordinate or dual momentum space from global mechanical ones.},
    \begin{align}
    &G(\Delta)=\frac1{4\sqrt{\sigma}}
    \Big[\,\frac{2w}{C\varOmega}\, H^2
    -\frac{N}{a^3}\frac{k}{\Delta + 3 k}\,\Big],        \label{coeff1}\\
    &V(\Delta)= \frac{2w}{C\varOmega} NaH(\Delta + 3 k), \\
    &U(\Delta)= 4 N a \sqrt{\sigma} (\Delta + 3 k)\Big[1+ \frac{2wNa}{C\varOmega}(\Delta + 3 k)\Big].     \label{coeff3}
    \end{align}
The above equations for $S_s^{(>0)}[\,\psi, \varPi_\psi]$ and $\mathcal H_*[\,\psi, \varPi_\psi]$ apply only in the spatially inhomogeneous sector of the theory. In the closed cosmological model with the discrete set of modes on the 3-sphere, $\psi(x)\to\psi_{n}$, this means that the space integral should be replaced as $\int d^3x\to\sum_{n=2}^\infty$, $G(\Delta)\to G(-n(n+2))$, etc. For spatially flat Friedmann background, $k=0$, the spectrum of degrees of freedom is continuous and can be represented by Fourier modes, $\psi(x)\to\hat\psi(p)$, $\int d^3x\to\int d^3p$, $G(\Delta)\to G(-p^2)$, which are adjacent at $p\to 0$ to the discrete homogeneous mode $\psi_{0}$.

\subsection{Perturbative (in)stability}
A nontrivial effect of solving the constraints is a complicated form of the coefficient $G(\Delta)$ in the kinetic term of (\ref{H_phys}) -- in contrast to the ghost mode in the homogeneous sector, its sign strongly depends on the details of the model. The positivity of this coefficient guarantees absence of ghost modes and conversion of the action to the Lagrangian form with the canonical normalization of the physical mode, $\psi\to\varphi$. Exclusion of the canonical momentum via its equation of motion, $\varPi_\psi=G^{-1}(\dot\psi-V\psi)$, gives the Lagrangian action
    \begin{align}
    &S_s^{(>0)}[\,\psi\,] =
    \int dt \,d^3 x\,\Big[\,\frac12\,\dot\psi\, G^{-1} \dot\psi
    - \dot\psi\,G^{-1}V\psi\nonumber\\
    &\qquad\qquad\qquad-\frac12\,\psi\,(U-VG^{-1}V)\psi\,\Big].
    \end{align}
For positive $G$ in terms of the canonically normalized field $\varphi$,
    \begin{equation}
    \psi=\sigma^{1/4}\sqrt{G}\,\varphi,
    \end{equation}
it reads after several integrations by parts
    \begin{align}
    &S_s^{(>0)}[\,\psi\,] =
    \frac12\,\int dt \,d^3 x\,\sqrt\sigma\,\left\{\,\dot\varphi^2
    - \varphi\,\Big[\,GU-V^2\right.\nonumber\\
    &\qquad\quad\left.+\mfrac12\,\partial_t\Big(\mfrac{\dot G}{G}\Big)-\mfrac14\Big(\mfrac{\dot G}{G}\Big)^2+V\partial_t\ln\mfrac{G}V\,\Big]\,\varphi\right\}.
    \end{align}

A remarkable property of the coefficients (\ref{coeff1})-(\ref{coeff3}) is that the combination $GU-V^2$ is linear in $\Delta$, so that non-polynomial in spatial derivatives (nonlocal) nature of the action takes place only for a closed model and is significant only for long wavelengths modes with a small $\Delta$,
    \begin{equation}
    G=\frac{wH^2}{2\sqrt\sigma\, C\varOmega}
    +O\Big(\mfrac{k}\Delta\Big).                   \label{G}
    \end{equation}
In this limit the potential part of the action represents the sum of gradient squared and the mass term with the effective mass parameter $m^2_{\rm eff}=m^2_{\rm eff}(a)$ which explicitly depends on $a$,
    \begin{eqnarray}
    &&GU-V^2+\mfrac12\,\partial_t\Big(\mfrac{\dot G}{G}\Big)-\mfrac14\Big(\mfrac{\dot G}{G}\Big)^2+V\partial_t\ln\mfrac{G}V\nonumber\\
    &&\qquad\quad=-c_s^2\frac\Delta{a^2}\,N^2
    +m^2_{\rm eff}N^2+O\Big(\mfrac{k}\Delta\Big),
    \end{eqnarray}
and $c_s$ -- the speed of sound parameter defined by
    \begin{equation}
    c_s^2=\frac{w(1+w)}\varOmega.     \label{sound}
    \end{equation}
Thus, the short wavelengths part of the action -- the limit of big $\Delta$ -- takes the form
    \begin{align}
    &S_s^{(>0)}[\,\psi\,] =
    \frac12\,\int dt \,d^3 x\,\sqrt\sigma\,
    \Big\{\,\dot\varphi^2\nonumber\\
    &\qquad\qquad\qquad\qquad+N^2\varphi\,
    \Big[\,c_s^2 \,\frac\Delta{a^2}
    -m^2_{\rm eff}\,\Big]\,\varphi\Big\}.         \label{action10}
    \end{align}
For spatially flat Friedmann model this representation is exact for the full range of spatial gradients $\Delta=-p^2$ -- the square of the comoving momentum.

Absence of ghost instability, $G>0$, and gradient instability, $c_s^2>0$, implies in view of (\ref{G}), (\ref{sound}) and the positivity of the constant $C$ (cf. Eq.(\ref{C})) the following two inequalities
    \begin{equation}
    \frac{w}\varOmega>0, \quad 1+w>0.     \label{stability_domain}
    \end{equation}
Outside of this range generic GUMG theory is unstable either due to ghost or gradient instability of the infinite set of modes extending to UV limit. Note that the second inequality is consistent with conclusions that violation of null energy condition is associated with perturbative instabilities \cite{Dubovsky_et_al,Sawicki_Vikman}.

Of course, there is always a ghost spatially homogeneous mode which, as was mentioned above, is just the perturbation of the minisuperspace Friedmann background. This mode can hardly disprove the GUMG theory, because this type of instability, or presence of a runaway solution, is just the essence of the cosmological acceleration or inflation phenomenon. A wrong sign of the kinetic term of the homogeneous background mode does not make general relativity an unstable theory, which is usually explained by the fact that this mode is not dynamically independent. In GUMG theory this mode is the physical one, and its effect is measured by the magnitude of the constant $C$ which is a part of freely specifiable initial conditions -- just like the cosmological constant as a constant of integration in UMG theory. Otherwise, it works dynamically exactly the same way as the scale factor mode in GR.

A possible objection to the stability of the GUMG model might be that simultaneous growth of negative energy background mode and the positive energy inhomogeneous modes is not prohibited by the conservation law for the total energy. But this objection is likely to be refuted by a simple argument that this is just a back reaction of cosmological perturbations on the Friedmann background, and the issue of this problem, or the magnitude of this effect, is far from being settled both at the classical and, even more so, at the quantum level \cite{Woodard, Garriga-Tanaka}. Thus, GUMG theory can be added to the list of reasons \cite{Smilga,Vikman} why in certain cases ghost modes are not harmful in physical models.

The linearized theory analysis above does not stand a smooth limit $w=-1$ which corresponds to the UMG theory, because of the singularity caused by $\varOmega=0$. This is because the number of local degrees of freedom drops in this limit to two, and the theory still remains stable even though it falls out of the domain (\ref{stability_domain}). Finally, below the phantom divide line $w=-1$ the extra scalar mode again becomes dynamical, but it suffers gradient instability with a negative $c_s^2$.

We will not discuss here other sources of instability, like the tachyonic one associated with the sign of the effective mass term in (\ref{action10}). The criteria of stability of the theory on a nonstationary background are very sophisticated when the positivity of the Hamiltonian, which is not conserved, becomes parametrization dependent and no longer indicative of the consistency of the model \cite{Sorokin_et_al}. In particular, competing contributions of time dependent comoving momentum and mass-like terms in (\ref{action10}) are subject to cosmological perturbation theory which reveals the particle creation and formation of inflationary power spectra. These phenomena will be considered elsewhere \cite{work_in_progress}, whereas below we will only briefly discuss the prospects of applying the GUMG model in context of dark energy and inflation theories.

\section{Discussion and conclusions}
Generalized unimodular gravity theory turned out to have a very rich dynamical structure. Its canonical formalism incorporates four generations of constraints of both the first and the second class, the first class ones being nontrivial linear combinations of the constraints belonging to different (primary and secondary) generations. Moreover, the number of primary first class constraints seemingly contradicts the originally claimed local symmetries of the Lagrangian action of GUMG model \cite{darkness}, which would indicate the breakdown of the Dirac conjecture in the theory of constrained dynamics. It turned out, however, that this conjecture still applies in GUMG model, because one of the originally assumed Lagrangian symmetries is actually either nonlocal in time or violates boundary conditions at spatial infinity. The constrained Hamiltonian formalism clearly reveals this peculiarity, which is deeply hidden in the Lagrangian framework, and allows one to recover true local symmetries as canonical transformations acting in the phase space of the theory.

Another peculiarity of this model is the so-called bifurcation of the system of constraints, which means that the theory has two branches with different numbers of constraints belonging to different classes. One branch is characterized by the set of first class constraints and can be interpreted as general relativity within a partial gauge fixation of spacetime diffeomorphisms, corresponding to the kinematical restriction on metric coefficients in GUMG model -- the lapse function as a rather generic function of the 3-metric determinant. The second, physically most interesting branch, is the one in which this restriction gives rise to the effective perfect fluid originally suggested as a candidate for dark energy. This dark fluid has a barotropic equation of state $p=w\varepsilon$ with a variable parameter $w=w(a)$ depending on the cosmological scale factor \cite{darkness}. This branch has only two local diffeomorphism symmetries realized as canonical transformations on phase space -- this enlarges the physical sector of the theory from two general relativistic degrees of freedom to three degrees of freedom. Perturbative analyses on the closed and spatially flat Friedmann background shows that this extra degree of freedom -- the scalar graviton -- can be free of ghost and gradient instabilities in a wide class of GUMG models satisfying the restrictions (\ref{stability_domain}). The second branch also includes a well-known unimodular gravity model corresponding to a constant value $w=-1$ which belongs to the boundary of this stability domain, but the theory is still stable, because the scalar mode is not dynamical one -- all the constraints of UMG model are the first class ones and they rule this mode out of the physical sector.

Even though GUMG theory was originally suggested as a model for cosmological acceleration, by and large it fails to accommodate the dark energy phenomenology. Cosmological data suggests \cite{DE_data,DE_data1} that the effective parameter $w$ could be below $-1$ at small $z$, which contradicts (\ref{stability_domain}). One of the possibilities to relax this stability criterion could be an attempt to construct a GUMG model in which the unstable scalar mode $\psi$ is not physical -- the method analogous to elimination of ghost modes by canonical constraints in \cite{Langlois_et_al}. Quite interestingly, it is possible.

If one chooses the function $w(a)$ in such a way that $\varOmega=0$, the sequence of canonical constraints (\ref{secondary})-(\ref{quat1}) does not terminate at the fourth generation, because the conservation of (\ref{quat1}) is no longer an equation on the Lagrange multiplier. In addition, two more constraints are generated, $\Delta(\Delta+3k)A=0$ and $\Delta B=0$. As a result all six scalar variables $(\psi,\varPi_\psi,E,\varPi_E,B,\varPi_B)$ are ruled out by the six constraints, and only harmless transverse-traceless tensor modes survive in the inhomogeneous sector of the model. In view of the expression (\ref{Omega}) for $\varOmega$ its zero value implies the following dependence of $w$ and $N$ on the scale factor
    \begin{equation}
    w = \frac{v}{a^3-v}, \quad N=n\,\frac{v-a^3}{va^3},
    \end{equation}
where $v$ and $n$ are some integration constants. Interesting case of $w<-1$ implies that $v>a^3$. In the course of cosmological expansion when $a$ grows to $a_0=v^{1/3}$ the barotropic parameter $w\to-\infty$ and $N\to 0$, so that $\dot a\to 0$ as it follows from the modified Friedmann equation (\ref{C}). The scale factor reaches maximum with the infinite {\em physical} Hubble parameter $\dot a/Na\sim (v-a^3)^{-1/2}$ and infinite acceleration in the cosmic (proper) time, $d^2a/d\tau^2=d^2a/(Ndt)^2\to\infty$. This is the analogue of the Big Rip singularity \cite{Big_Rip} beyond which the physical evolution cannot be analytically continued. For a negative $v$ the parameter $w$ stays above its value $-1$ and smoothly grows to zero in the course of expansion, which is again different from the dark energy scenario, because it does not cross the phantom divide line $w=-1$ and resembles more the exit from inflation picture. Altogether, including this case of nondynamical scalar graviton, it seems that GUMG model is more interesting as a new candidate for inflation scenario, rather than the source of dark energy.

The GUMG model of inflation could be rather interesting, because the role of the inflaton is played by the scalar sector of the metric field. The situation is similar to the $R^2$-inflation in the Starobinsky model \cite{Starobinsky_model}, when the role of inflaton is played by the dynamical conformal mode of the metric. Here the same is realized without higher order derivatives of the metric -- entirely due to kinematical restriction on the lapse function of the theory. In view of the expression (\ref{Omega}) for $\varOmega$, stability conditions (\ref{stability_domain}) for the scalar graviton at negative $w$ imply that
    \begin{equation}
    \frac{dw}{d\ln a}>-3w(1+w)>0,     \label{stable1}
    \end{equation}
which is consistent with the inflation scenario, because $w$ grows to zero at the exit from inflation, just like in the zero $\varOmega$ case above.

Moreover, the scalar graviton $\psi$ with the action (\ref{action10}) has a dispersion equation with a nontrivial speed of sound (\ref{sound}) determined by the time-dependent equation of state of the effective perfect fluid $p=w(a)\,\varepsilon$. It is interesting that in the hydrodynamical formalism of inflation theories, incorporating also the class of k-essence models \cite{Mukhanov}, the speed of sound (critically effecting the primordial power spectrum of perturbations) is determined by the effective equation of state of the inflaton field, $c_s^2=dp/d\varepsilon$. In the GUMG model this expression equals
    \begin{equation}
    \frac{dp}{d\varepsilon} =w+\varepsilon\frac{dw/da}{d\varepsilon/da}=
    w-\frac13\,\frac1{1+w}\,\frac{dw}{d\ln a},
    \end{equation}
where we took into account the stress tensor conservation law (\ref{epsilon_equation}) for $\varepsilon$. On the other hand, in view of the expression (\ref{Omega}) for $\varOmega$, the sound of speed of the GUMG scalar graviton (\ref{sound}) is the same up to quadratic order in $dw/da$
    \begin{equation}
    c_s^2=\frac{w}{1+\frac1{3w(1+w)}\,\frac{dw}{d\ln a}}=\frac{dp}{d\varepsilon}
    +O\left(\,\big(dw/da\big)^2\,\right).
    \end{equation}
For slowly varying functions $w(a)$ these two parameters nearly coincide. For faster varying equations of state their discrepancy is not surprising, though, because the dynamics of the usual inflaton field additional to gravity is qualitatively different from the case when the inflaton belongs to the metric sector and acquires a dynamical nature due to nontrivial GUMG kinematics.

Thus, despite the fact that generalized unimodular gravity most likely represents a failed attempt to build a phenomenologically consistent model of dark energy, this theory can be regarded as a rather prospective source of inflation scenario, which is a subject of further studies \cite{work_in_progress}.

\section*{Acknowledgments} We want to express special thanks to Alexander Kamenshchik with whom the idea of the generalized unimodular gravity model was originally put forward. Also we are indebted to Joseph Buchbinder, Mario Herrero-Valea, Nobu Ohta, Sergey Sibiryakov, Dima Sorokin, Philip Stamp, Gian Paulo Vacca and Alex Vikman for stimulating discussions. One of the authors (A.B.) is grateful for hospitality of the Yukawa Institute for Theoretical Physics, Pacific Institute for Theoretical Physics and Green College of the University of British Columbia, where part of this work has been completed. This work was supported by the RFBR grant No.17-02-00651.

\end{document}